\documentclass[iop]{emulateapj}
\usepackage{lineno,xcolor,hyperref}

\newcommand\lbol{$L_{\rm bol}$}
\newcommand\msun{$M_{\odot}$}
\newcommand\lsun{$L_{\odot}$}
\newcommand\logl{$\log (L/L_\odot)$}
\newcommand\um{$\mu$m}

\def\rsgage{$10.4^{+1.3}_{-1.2}$Myr}

\shorttitle{The age of Wd1}
\shortauthors{Beasor et al.}
\graphicspath{{./}{figures/}}

\begin{document}

\title{The age of Westerlund 1 revisited}

\email{embeasor@gmail.com}

\author{Emma R. Beasor}\altaffiliation{Hubble Fellow}
\affiliation{NSF's OIR Lab \\
950 N. Cherry Ave., Tucson, AZ 85721, USA}
\author{Ben Davies}
\affiliation{Astrophysics Research Institute, Liverpool John Moores University \\
146 Brownlow Hill, Liverpool, L3 5RF, UK}
\author{Nathan Smith}
\affiliation{Steward Observatory, University of Arizona \\
933 N. Cherry Ave., Tucson, AZ 85721, USA}
\author{Robert D. Gehrz}
\affiliation{Minnesota Institute for Astrophysics, School of Physics and Astronomy \\
116 Church Street SE, University of Minnesota, Minneapolis, MN 55455, USA}
\author{Donald F. Figer}
\affiliation{Rochester Institute of Technology \\
54 Memorial Drive, Rochester, NY 14623, USA}




\begin{abstract}

The cluster Westerlund~1 (Wd1) is host to a large variety of post main-sequence (MS) massive stars. The simultaneous presence of these stars can only be explained by stellar models if the cluster has a finely-tuned age of 4-5Myr, with several published studies independently claiming ages within this range. At this age, stellar models predict that the cool supergiants (CSGs) should have luminosities of $\log(L/L_\odot) \approx 5.5$, close to the empirical luminosity limit. Here, we test that prediction using archival data and new photometry from SOFIA to estimate bolometric luminosities for the CSGs. We find that these stars are on average 0.4dex too faint to be 5Myr old, regardless of which stellar evolution model is used, and instead are indicative of a much older age of $10.4^{+1.3}_{-1.2}$Myr. We argue that neither systematic uncertainties in the extinction law nor stellar variability can explain this discrepancy. In reviewing various independent age estimates of Wd1 in the literature, we firstly show that those based on stellar diversity are unreliable. Secondly, we re-analyse Wd1's pre-MS stars employing the Damineli extinction law, finding an age of $7.2^{+1.1}_{-2.3}$Myr; older than that of previous studies, but which is vulnerable to systematic errors that could push the age close to 10Myr. However, there remains significant tension between the CSG age and that inferred from the eclipsing binary W13. We conclude that stellar evolution models cannot explain Wd1 under the single age paradigm. Instead, we propose that the stars in the Wd1 region formed over a period of several Myr.  

\end{abstract}



\section{Introduction}

Westerlund 1 (Wd1) is a highly extinguished cluster located in the Milky Way, first discovered by \citet{westerlund1961population}.  We now know that Wd1 is host to a large number of both hot and cool evolved massive stars, including Wolf-Rayet stars \citep{crowther2006census}, a luminous blue variable \citep{ritchie2009lbv}, late-type supergiants (red and yellow supergiants), as well as a magnetar \citep{muno2007magnetar}. Assuming this cluster consists of a simple stellar population (SSP, i.e. single age, single star evolution), there is a narrow range of ages for which all of these objects could coexist in a single cluster \citep[$\sim$4--5 Myr,][]{clark2005massive}. Indeed, there are numerous works in the literature that claim an age of 5Myr for Wd1 from various methods. These methods include fitting the number ratios of stars of different spectral types \citep{crowther2006census,dorn2018stellar}, isochrone fitting to pre main-sequence (MS) stars \citep{brandner2008intermediate,gennaro2011mass,kudryavtseva2012instant}, and from dynamical masses of stars in eclipsing binary systems \citep{ritchie2010vlt,koumpia2012fundamental}. 

 \begin{figure*}
    \centering
    \includegraphics[width=\textwidth]{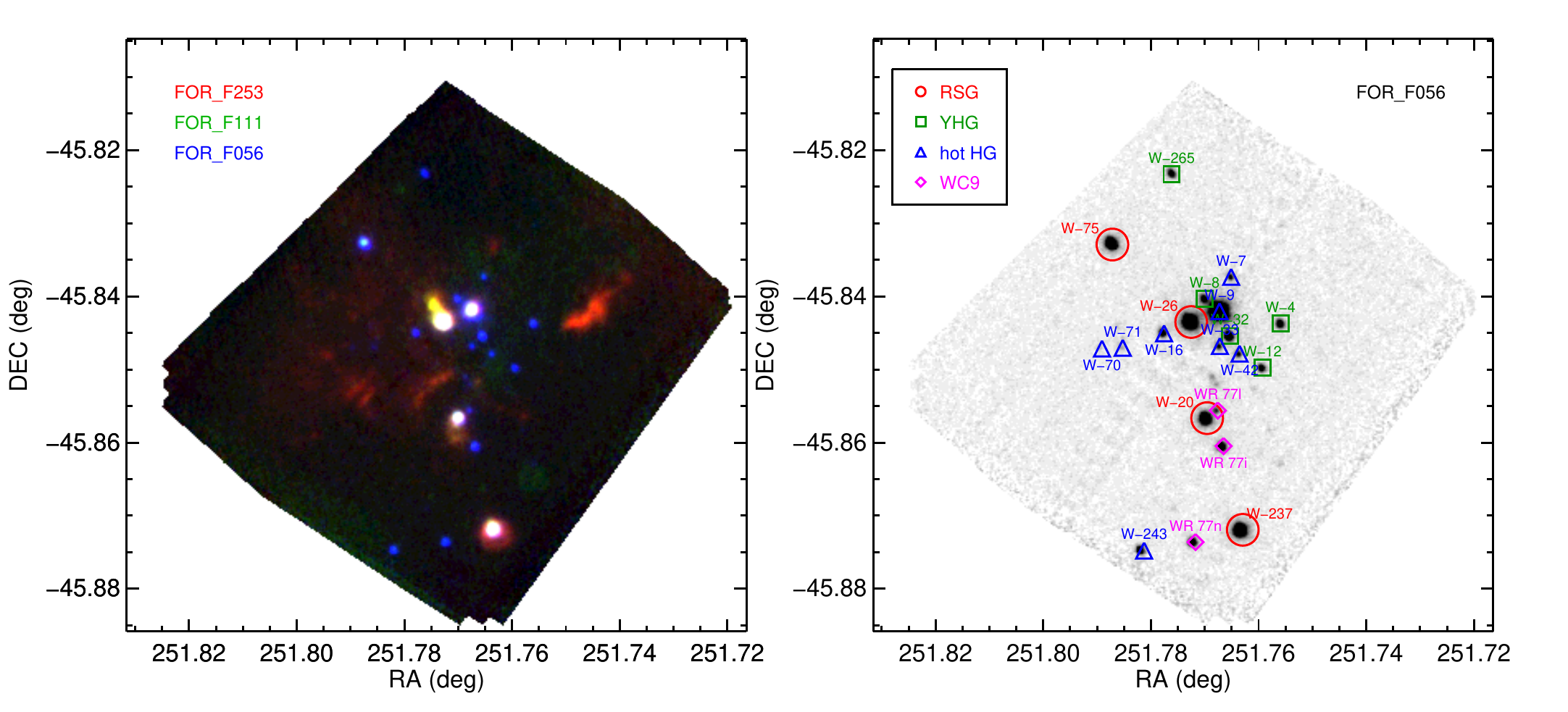}
    \caption{{\it Left panel:} 3-colour image of Wd1 using 5.6$\micron$, 11.1$\micron$ and 25.3$\micron$ filters.{\it Right panel:} Finding chart for Wd1.}
    \label{fig:3colour}
\end{figure*}

At an age of 5Myr, the expectation from all stellar evolution models is that the brightnesses of the YSGs and RSGs in Wd1 should be $\log(L/L_\odot) \ga 5.5$ \citep[see e.g. Fig.\ 5 in][]{ritchie2010vlt}, which corresponds to the empirically-derived luminosity limit for cool supergiants known as the Humphreys-Davidson (H-D) limit \citep{humphreys1979studies,davies2018humphreys}. However, to date this prediction has not been verified. Doing so requires accurate photometry in the near and mid-IR, but the Wd1 RSGs in particular are badly saturated in IR survey images of e.g. 2MASS, WISE and GLIMPSE, making bolometric luminosity estimates of these stars difficult. 

In this paper, we present new SOFIA-FORCAST photometry for Wd1, covering 5.6 - 31.5$\micron$, which we combine with shorter wavelength photometry from the literature, and determine the first independent measurements of \lbol\ for the RSGs and YSGs in Wd1 (Sect.\ \ref{sec:csglbol}). In Sect.\ \ref{sec:hr} we compare these luminosities to the expectations for a 5Myr old stellar population. Finally, in Sect.\ \ref{sec:disco} we take a critical look at previous investigations into Wd1's age, to try and reconcile those with the cool supergiant luminosities.

\section{Cool supergiant luminosities} \label{sec:csglbol}
\subsection{Data}\label{sec:data}
We obtained new mid-IR data using FORCAST \citep{herter2012first} on SOFIA \citep[][Prog ID: 05 0064, PI: N Smith]{young2012early} in Cycle 5, observing the cluster in filters 5.5, 7.7, 11.1, 25.3 and 31.5$\mu$m. These wavelengths cover a crucial part of the spectral energy distribution (SED) that cannot be observed from the ground. We list the exposure times and observation dates for each filter in Table \ref{tab:exptimes}. All observations were taken in the asymmetric chop with offset nod (C2NC2) mode.  The data were reduced using SOFIA data pipeline {\tt FORCAST Redux} \citep{Clarke2015}, and in this work we use Level 3 flux-calibrated data. As the field is not crowded in our data (see Fig. \ref{fig:3colour}) we use aperture photometry to extract the stars' fluxes. Aperture sizes were defined manually so as to encompass all flux above the sky noise level at each wavelength. To assess the level of uncertainty introduced by aperture selection, we experimented with varying the aperture sizes for each star. We found that the results were stable at the 5\% level. This uncertainty was then added to the photometric errors in quadrature.

\begin{figure*}
    \includegraphics[width=0.49\textwidth]{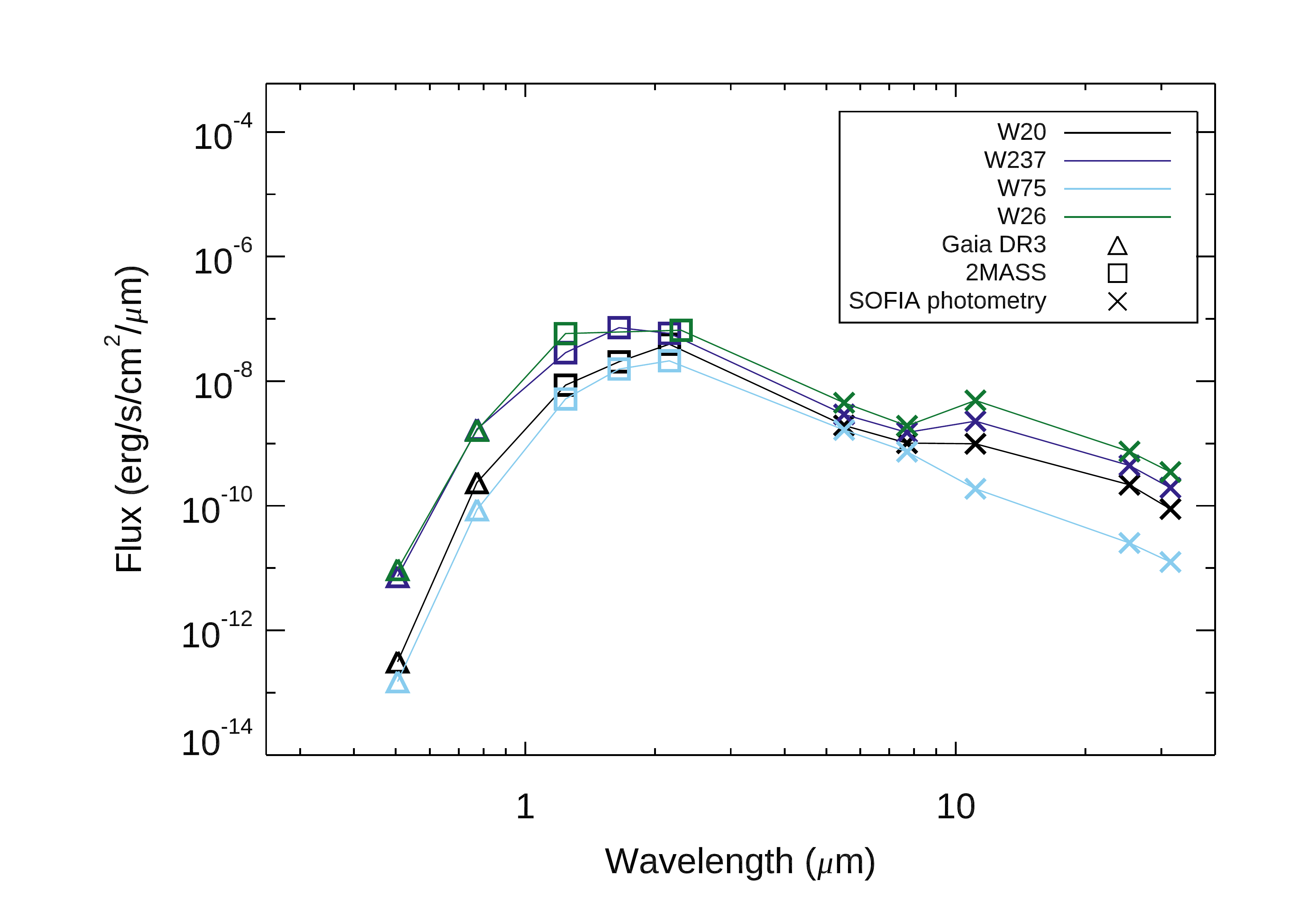} 
    \includegraphics[width=0.49\textwidth]{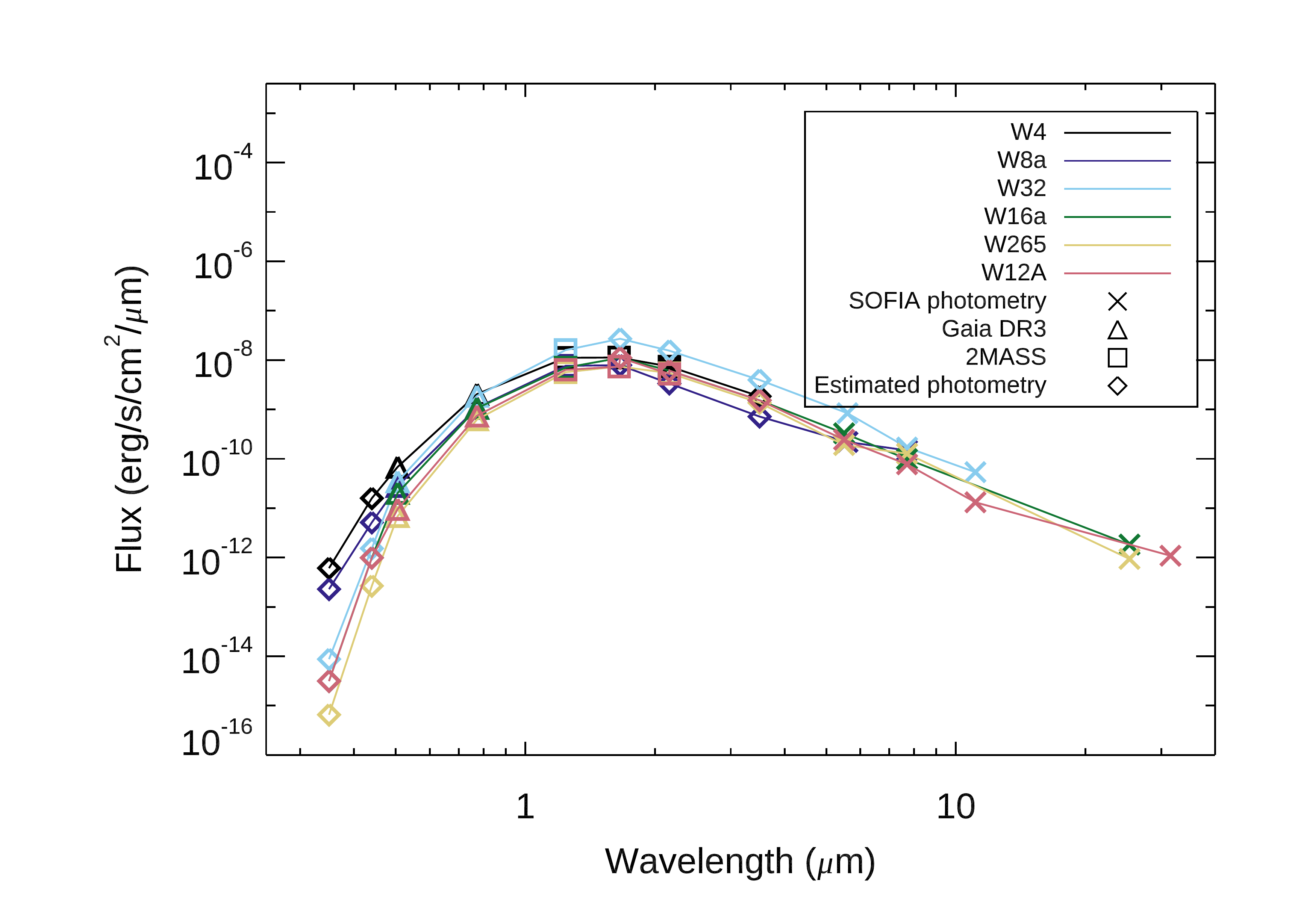} 
    \caption{{\it Left: }SEDs showing available photometry for the four RSGs in Westerlund 1. {\it Right: } SEDs showing available photometry for the YSGs, as well as estimated photometry indicated by the triangle symbol. Formal photometric errors are smaller than the plotting symbols.}  
    \label{fig:photometry}
\end{figure*}

Combining these new observations with archival photometry from Gaia EDR3 (Gaia Collaboration, in prep, 2020b) and 2MASS \citep{skrutskie2006two2} we now have a full bolometric SED for each of the RSGs in Wd1, shown in Fig. \ref{fig:photometry}\footnote{While these objects have been imaged in other infrared missions (e.g. IRAC and MSX \citep{fazio2004irac,price2001midcourse}) the photometry was often saturated or provided upper limits only. For this reason, we utilise only our new SOFIA-FORCAST observations to cover the mid-IR region of the SED. }. For most CSGs in Wd1, the 2MASS photometry is somewhat uncertain due to the stars being close to the saturation limit, resulting in photometric errors of up to 40\%. For W26, there was no usable H or K photometry from 2MASS at all. To cover this part of the SED for W26, we take the 2.3$\mu$m photometry from \citet{mengel2007medium}. 

\begin{table}
    \centering
       \caption{SOFIA-FORCAST observation details. }
    \begin{tabular}{lccc}
    \hline
         Filter ($\mu$m) & Date observed  &Exposure time (s)  \\
         \hline
         5.6 & 6th August 2017 &111.7\\
         7.7 & 6th August 2017& 599.5\\
         11.1 &6th August 2017 & 209.2\\
         25.4 & 3rd August 2017 & 617.8 \\
         31.5 & 3rd August 2017 &654.0\\
         \hline
    \end{tabular}
 
    \label{tab:exptimes}
\end{table}

\subsection{Distance}\label{section:wd1properties}
There has been a large amount of conjecture over the distance to Wd1, discussed at length by \citep{aghakhankoo2019inferring}. Most previous studies have adopted a distance of $\sim$4 -- 5 kpc \citep[e.g.][]{crowther2006census,clark2005massive}. \citeauthor{aghakhankoo2019inferring} themselves estimate a distance to Wd1, using the Gaia DR2 parallax distribution of all stars in the field of Wd1. These authors model the stars in their sample as being a superposition of two components, cluster and field, and arrive at a distance for the `cluster' component of 2.6$^{+0.6}_{-0.4}$kpc, significantly lower than previous estimates. 

Most recently, \citet[][hereafter DB19]{davies2019distances} estimated the distance to Wd1 also using Gaia DR2 data. In this work the sample of stars used to find the average cluster parallax are chosen based on OB spectral types and whether or not their proper motions (PMs) are consistent with the cluster average. While this leaves a smaller sample of stars than that of \citet{aghakhankoo2019inferring}, the likelihood of genuine cluster membership of the remaining stars is extremely high. DB19 find a distance of 3.87$^{+0.95}_{-0.64}$kpc, in excellent agreement with the distance from \citet{Rate2020unlocking} which employed a similar methodology and Gaia DR2. 

In this work, we use the distance obtained via the same method as in DB19, but with the updated Gaia-EDR3 astrometry \citep{GaiaDR3}. In this data release, as well as an improvement in astrometric precision, there is also a better understanding of Gaia's zero-point parallax offset \citep{Lindegren20}. We select stars with known OB spectral types within 5\arcmin\ of the cluster centre, excluding known binaries. We also exclude stars whose proper motions are outside $\pm$10$\rm km\,s^{-1}$ of the cluster average, this limit being the approximate virial velocity dispersion for a cluster of Wd1's mass ($\sim 10^5$\msun) and size ($\sim$1pc). The parallaxes ($\pi$) of the individual stars are corrected for the zero-point offset according to their position on the sky, brightness and colour\footnote{The stars in Wd1 are very red owing to the large amount of foreground extinction, which puts the stars right at the edge of the calibrated parameter space \citep{Lindegren20}. Though we apply the zero-point calibration prescription according to each star's brightness, colour and position, we note that the uncertainty in the zero-point remains the major source of uncertainty, as in DB19.}. The probability distribution for the average parallax $\bar{\pi}$ is determined by combining those for the individual stars. The error on $\bar{\pi}$ is found via Monte-Carlo trials, in which we repeat the above but with parallaxes randomly sampled from each star's distribution. The average parallax we find for Wd1 is $\bar{\pi} = 0.243 \pm 0.027$, which corresponds to a distance of $d = 4.12^{+0.66}_{-0.33}$kpc. This is slightly further than the Gaia DR2 distances of \citet{davies2019distances} and \citet{Rate2020unlocking}, but consistent to within the errors.

\begin{table*}
\centering
\caption{Photometry for Westerlund 1 CSGs from SOFIA-FORCAST. All photometry is in Jy. }
\label{table:wd1photom}
\begin{tabular}{lcccccc}
\hline\hline
ID &5.5$\mu$m &7.7$\mu$m &11.1$\mu$m& 25.3$\mu$m &  31.5$\mu$m \\ [0.5ex] 
\hline
W237&    29 $\pm$ 1& 30 $\pm$ 2 &93.7$\pm$ 1.6 &94.6 $\pm$ 1.6 & 64.33 $\pm$ 0.7 \\
 W20&    19.6$\pm$ 1.9 & 20.0 $\pm$ 0.42&40.6 $\pm$ 2.0& 46.4$\pm$ 5.2& 29.4$\pm$ 5.4 \\
 W26&    45.38 $\pm$ 4.2& 38$\pm$ 3&202 $\pm$ 6&159$\pm$ 14&115 $\pm$ 11 \\
 W75&    16.6 $\pm$ 0.6& 14.6 $\pm$ 0.8& 7.7$\pm$ 0.7& 5.4 $\pm$ 0.7& 4.1 $\pm$ 1.1 \\
 \hline
  W4&    3.82$\pm$ 0.03& 2.40 $\pm$ 0.03& 1.03 $\pm$ 0.08& 0.8 $\pm$ 0.1& 0.25 $\pm$ 0.09 \\
 W8a&    2.29 $\pm$ 0.02 & 2.92 $\pm$ 0.06 &-&-&- \\
 W32&    8.76 $\pm$ 0.09& 3.34$\pm$ 0.02& 2.2 $\pm$ 0.2&-& - \\
W16A&    3.31 $\pm$ 0.07& 1.95 $\pm$ 0.03&-& 0.39 $\pm$ 0.22 & - \\
W265&    3.08 $\pm$ 0.03& 1.74 $\pm$ 0.02& -& 0.2 $\pm$ 0.1& - \\
W12A&    2.46 $\pm$ 0.02&1.53 $\pm$ 0.03&0.54 $\pm $ 0.1 &- & 0.4 $\pm$ 0.1 \\
\hline
\end{tabular}
\end{table*}

\begin{table*}
\centering
\caption{Archival photometry for Westerlund 1 CSGs.}
\label{table:wd1photom_arch}
\begin{tabular}{lccccccc}
\hline\hline
ID & 2MASS-J (Jy) & 2MASS-H (Jy) & 2MASS-K$_{\rm s}$ (Jy)  & $K_{\rm s}$ (M\&T07)& Gaia-BP (mJy) & Gaia-RP (mJy)  \\ [0.5ex] 
&  1.24 $\mu$m& 1.65 $\mu$m& 2.16 $\mu$m &2.3 $\mu$m&0.505 $\mu$m& 0.772 $\mu$m \\ [0.5ex]
\hline
%
W237&   14.70 $\pm$  0.20 & 65.40 $\pm$ 16.20 & 90.90 $\pm$ 24.00 &- &0.624 $\pm$ 0.016 & 338.0 $\pm$ 4.0 \\
 W20&    4.43 $\pm$  0.04 & 18.60 $\pm$  1.30 & 61.00 $\pm$ 23.50 &- &0.027 $\pm$ 0.002 &  46.9 $\pm$ 0.4 \\
 W26&   29.80 $\pm$ 14.80 & -                 & -                 &116 $\pm$ 11.6 &0.807 $\pm$ 0.04 & 325.0 $\pm$ 4.0 \\
 W75&    2.66 $\pm$  0.05 & 14.20 $\pm$  0.30 & 33.00 $\pm$  8.30 &- &0.013 $\pm$ 0.002 &  17.0 $\pm$ 0.1 \\
 \hline
  W4&    5.70 $\pm$  0.12 & 10.20 $\pm$  0.20 & 11.50 $\pm$  0.20 &- &5.87 $\pm$ 0.06 & 406.0 $\pm$ 2.0 \\
 W8a&    3.91 $\pm$  0.07 & -                 & -                 & -&2.27 $\pm$ 0.02 & 206.0 $\pm$ 1.0 \\
 W32&    8.17 $\pm$  0.15 & -                 & -                 & -&2.90 $\pm$ 0.03 & 357.0 $\pm$ 2.0 \\
W16a&    3.60 $\pm$  0.07 & -                 & -                 & -&1.72 $\pm$ 0.02 & 206.0 $\pm$ 1.0 \\
W265&    2.91 $\pm$  0.05 & 6.72 $\pm$  0.19  & 8.48 $\pm$  0.13  & -&0.582 $\pm$ 0.007 & 119.0 $\pm$ 1.0 \\
W12A&    3.25 $\pm$  0.05 & 6.69 $\pm$  0.13  & 8.15 $\pm$  0.14  &- &0.776 $\pm$ 0.011 &  138.0 $\pm$ 1.0 \\

\hline
\end{tabular}
\end{table*}

\subsection{Luminosities}\label{section:rsglbols}
To obtain luminosities for the cool supergiants in Wd1, we integrate under the spectral energy distribution (SED), as in \cite{davies2018humphreys}. By doing so, we make the implicit assumption that any flux absorbed by circumstellar material (CSM) will be either scattered at a similar wavelength or thermally re-radiated by the CSM. By integrating under the entire SED we will therefore retrieve all of the flux from the star. This assumption holds true as long as the CSM is neither optically thick nor highly aspherical. In a study of hundreds of RSGs in the Large and Small Magellanic Clouds, \citet{davies2018humphreys} found only one star (WOH G64) for which this assumption appeared to break down. This star is so heavily embedded in its CSM that it is barely detectable in the optical, despite very little foreground extinction, whilst it is one of the brightest objects in the LMC at mid-IR wavelengths. Though all RSGs in Wd1 have some mid-IR excess, the excess emission is only a small fraction of the total flux ($\la$10\%). Hence, it seems unlikely these stars possess a large amount of CSM dust, possibly due to UV radiation from nearby O stars \citep{clark2019census}. For these reasons, we conclude that asphericity in the dust geometry of the RSGs is only a minor consideration in calculating their bolometric luminosities. Should the emission from the neighbouring hot stars indirectly pollute the RSG SEDs, this would cause us to overestimate the stars' luminosities, though we expect this effect to be minor.  


Before integrating under the SEDs, we must first correct for foreground extinction. The extinction towards Wd1 has also been discussed many times in the literature, spanning a large range \citep[A$_{\rm K}$ = 0.7 - 1.1, e.g. ][]{clark2005massive,crowther2006census,brandner2008intermediate,koumpia2012fundamental,damineli2016extinction}. However, more recent work seems to be reaching a consensus on the true extinction value. \citet{andersen2017very}, which supercedes similar work by the same group \citep{brandner2008intermediate,gennaro2011mass}, focuses on the colours of the low mass stellar content of Wd1, to determine an extinction value of $A_{\rm K}$ = 0.87$\pm$0.14. The state-of-the-art in this respect is \citet{damineli2016extinction}, who determined the optical/infrared extinction law for Wd1 from the colours of the hot stars, finding a much steeper power law than earlier studies \citep[e.g. ][]{clark2005massive}, and an average $A_{\rm K}$ of 0.74$\pm$0.08. Throughout this work we will adopt the extinction law of \citet{damineli2016extinction}, though we will explore the possible effect of systematic errors arising from the extinction law in Sect. \ref{sec:hr_extinct}.

For the Wd1 CSGs, we took all of the available photometry, dereddened it, interpolated between the photometric points on a logarithmic wavelength scale, and integrated underneath the SED using the IDL routine {\tt int$\_$tabulated}. We used a black body with temperature 3500K to estimate the flux emitted by each star at wavelengths shorter than V. Though this is a crude approximation of the SED shape below $\sim$0.5\um, our tests with model MARCS atmospheres show that $<10$\% of the star's flux is emitted at these wavelengths. This is clearly seen from a visual inspection of RSG spectrophotometry \citep[e.g.][]{davies2013temperatures}. Furthermore, these tests revealed that from the wavelength sampling of our photometry, combined with a Planck-function of temperature 3500-4500K to approximate the blue region of the spectrum we can obtain bolometric luminosities accurate to within 0.02dex. \footnote{To investigate the accuracy of our method, we performed a similar analysis on MARCS model SEDs \citep{gustafsson2008grid} with $T_{\rm eff}$ values between 3400-4500K. We convolve these model SEDs with filter profiles for VRIJHK and FORCAST photometry, again using a black body to estimate the flux emitted at wavelengths shorter than V. We then interpolated between the filter wavelengths in log space and integrated underneath the points to get a bolometric luminosity. We found that we were able to retrieve bolometric luminosities for these model SEDs to within 0.01 dex.  }

\begin{figure*}
    \centering
    \includegraphics[width=0.49\textwidth]{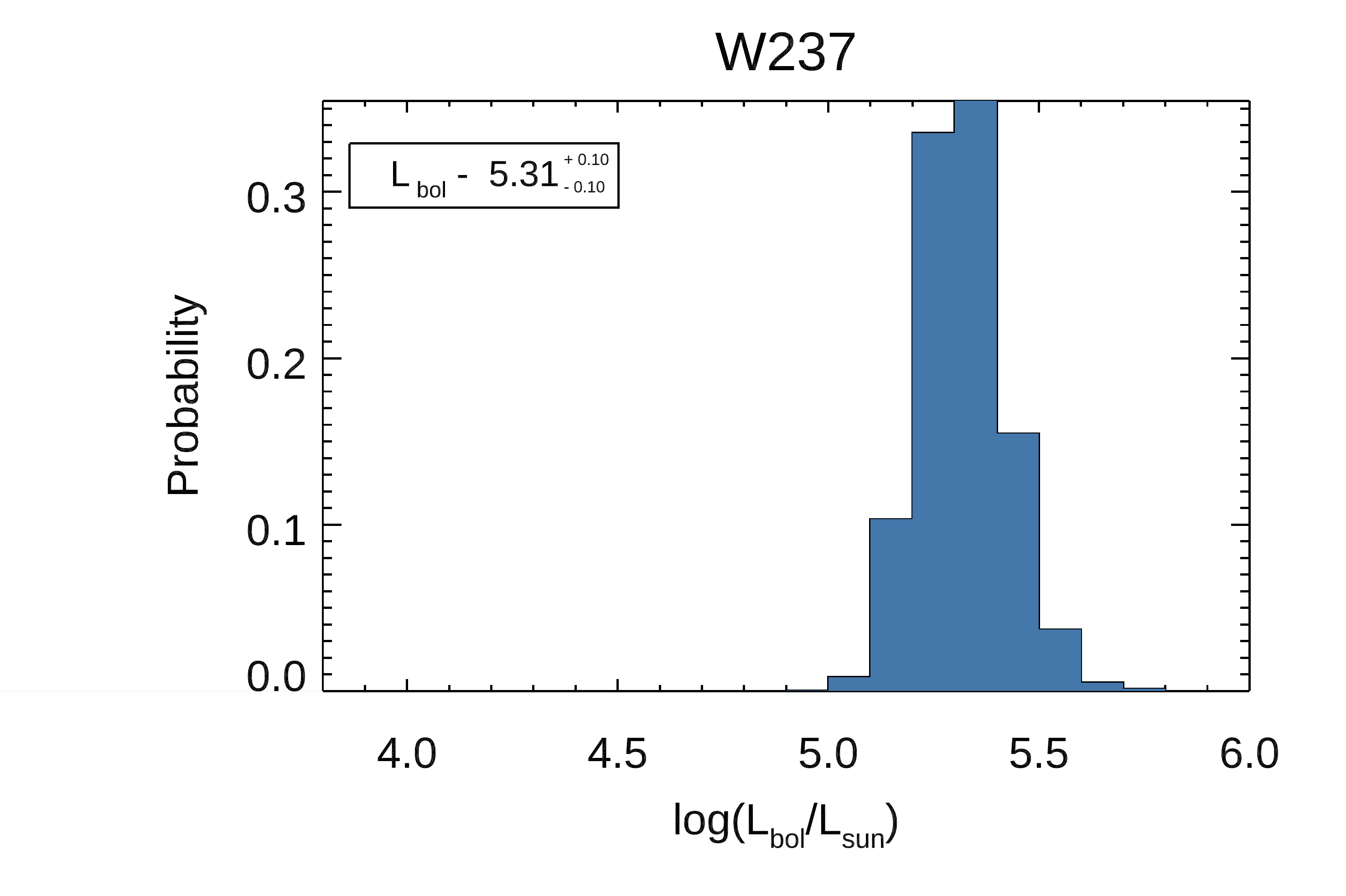}
    \includegraphics[width=0.49\textwidth]{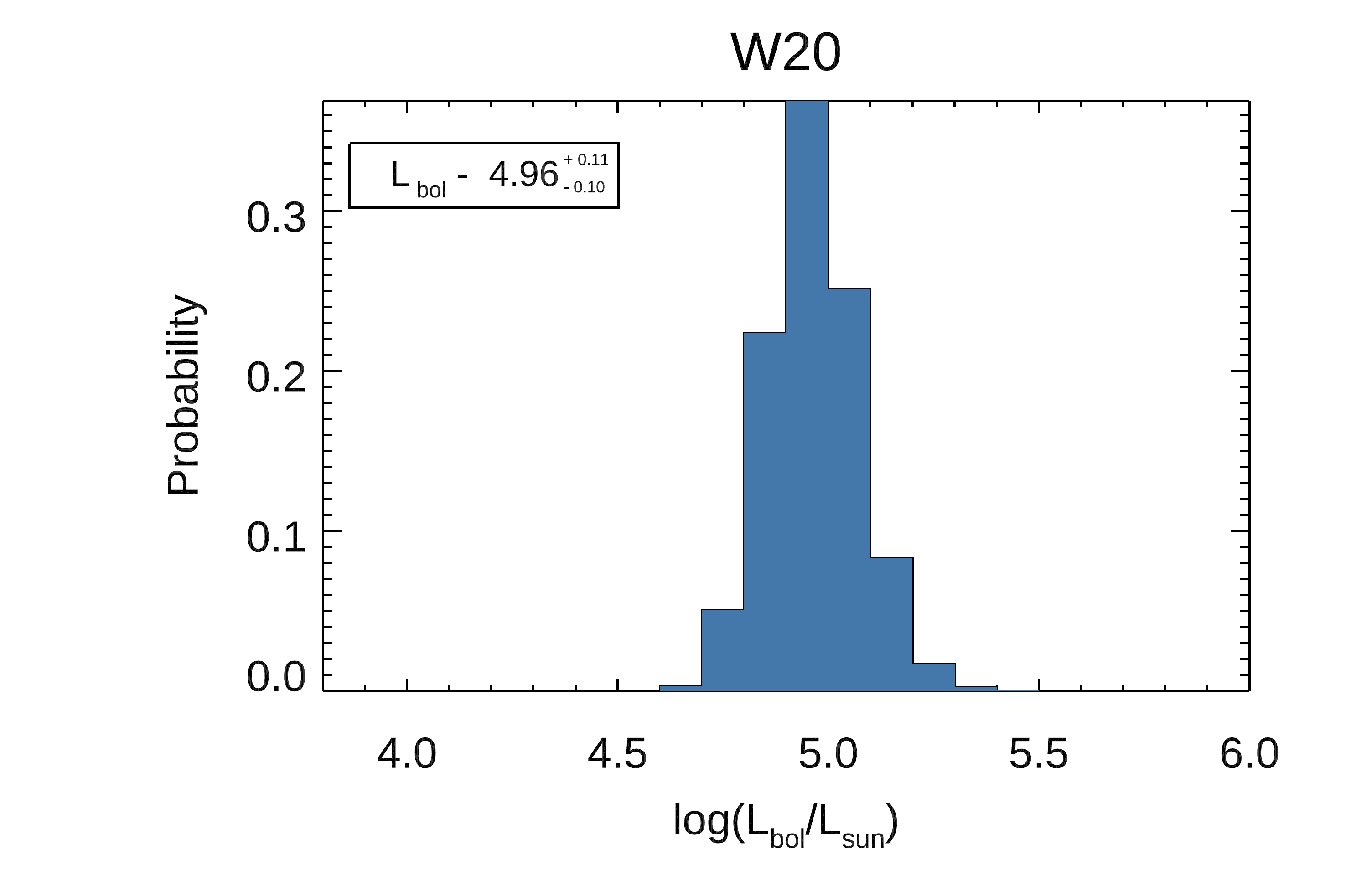}
    \includegraphics[width=0.49\textwidth]{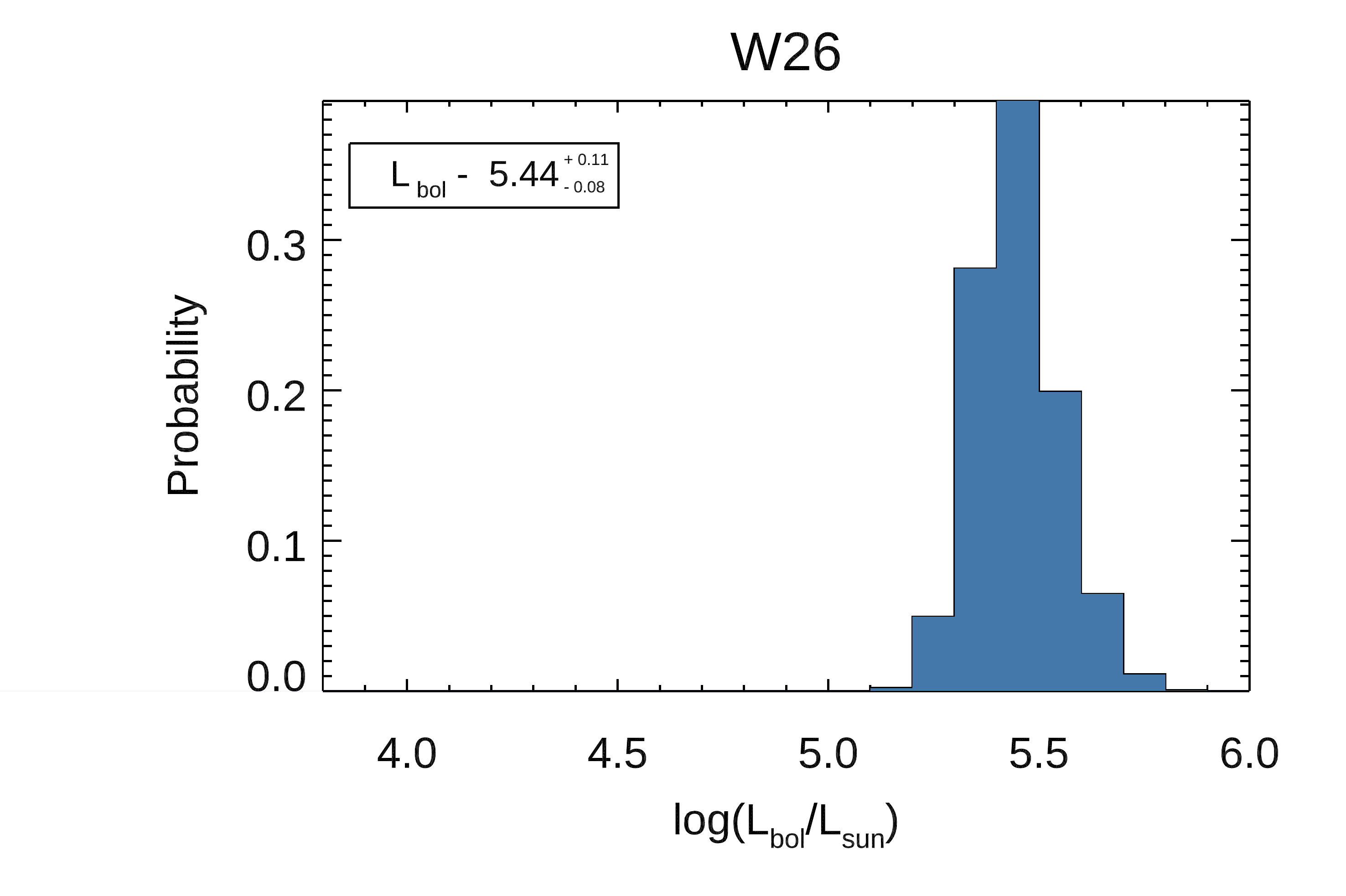}
    \includegraphics[width=0.49\textwidth]{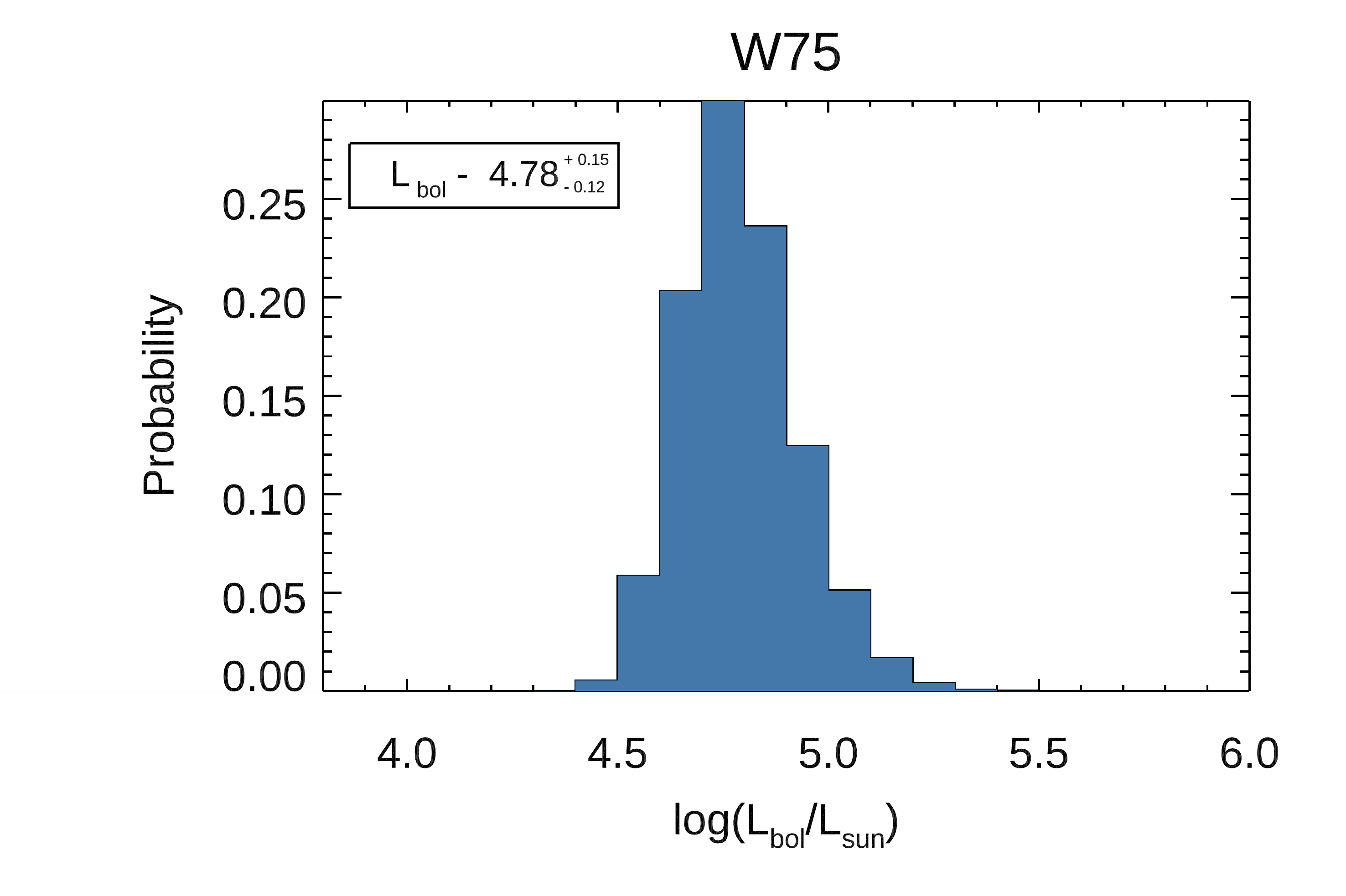}
    \caption{The luminosity probability distributions obtained for each RSG in the cluster. }
    \label{fig:lbol_dists}
\end{figure*}

To determine the errors on the luminosities, we consider the uncertainties in the extinction, distance, and the photometry itself. Of these, two have non-Gaussian probabilities, and so the errors cannot simply be added in quadrature. Instead, to determine the posterior probability distribution on \lbol\ for each star, we employ the following Monte-Carlo method. First, for each star we randomly sample the extinction from a Gaussian distribution centred on $A_{\rm K}$ = 0.74 \citep{damineli2016extinction} with a standard deviation of 0.08 mags, and deredden the observed SED using the corresponding extinction law. If the extinction value of a given trial results in the dereddened RSG [V-K] and [V-I] colours being too blue for the star's spectral type, we discard the results of that trial. Since the Wd~1 RSGs are likely to be spectroscopically variable, we use a conservative spectral type threshold of M0, which is several subtypes earlier than those listed by \citeauthor{clark2010serendipitous}. This process is repeated $10^4$ times, which allows us to build a posterior probability distribution for the luminosity of each RSG in Wd1, which we show in Fig.\ \ref{fig:lbol_dists}. We take the {\bf median} of the distribution as the most likely luminosity, and determine the uncertainties from the 68\% confidence intervals, analogous to the 1-sigma errors. These errors are dominated by those in distance and foreground extinction. 

 
For completeness, we also estimate the luminosities for each RSG using bolometric corrections to the $K_{\rm s}$ band photometry. Since all RSGs in Wd1 are of late-M spectral types, we adopt the bolometric correction ${\rm BC}_K =3.00\pm0.18$ appropriate for such stars \citep{davies2018initial}. For all RSGs except W26, the photometry comes from 2MASS, and as such suffers from saturation effects -- all have quality flag `D' in the 2MASS point-source catalogue and have photometric errors of 0.25-0.40mag. For W26, there is no available 2MASS data and so we instead take the $k_{\rm s}$-band flux from \citet{mengel2007medium}. We find all BC-based luminosities are consistent to within the errors of the SED method (see Table \ref{tab:lbols}), though we note that the BC-based luminosities of W237, W20 and W75 should be treated with caution due to the poor quality photometry.    


\subsubsection{Yellow supergiants}\label{section:ysglbols}
The luminosities for the YSGs in Westerlund 1 are estimated following a similar method to that for the RSGs. However, due to the higher effective temperature of YSGs, a larger proportion of their flux is emitted at wavelengths shorter than $V$-band. We therefore need to take an extra step to ensure \lbol\ is not underestimated.

First, we deredden the available photometry with the $A_{\rm K}$ value for the current trial using the \citet{damineli2016extinction} extinction law. For W8, W16 and W32, we then take the intrinsic colour corrections from \citet{koornneef1982gas} and \citet{fitzgerald1970intrinsic} and create synthetic UBHKL photometry (for the stars' spectral types, see Table \ref{tab:lbols_ysgs_tab}). For W4, W265 and W12, we only create synthetic $U$, $B$ and $L$ band photometry. Next, we again use a blackbody to account for any missing flux shorter than U-band (0.35$\micron$), assuming a $T_{\rm eff}$ appropriate for the spectral types of the YSGs listed in Table \ref{table:wd1photom_arch}. We again estimated systematic errors on the luminosity from the temperature of the blackbody by varying $T_{\rm eff}$ by $\pm$2000K, corresponding to a change in \lbol\ of $\pm$ 0.05 dex. We then integrate under this total SED to determine a value for \lbol\ for the current $A_{\rm K}$ and distance. We then use colour cuts to exclude any \lbol\ values from SEDs that are too blue for the stars' spectral types, as listed by \citet{koornneef1983near}. Our YSG results are presented in Table \ref{table:wd1photom_arch}. As with the RSGs, the uncertainties on\lbol\ are dominated by those on distance and extinction.

The YSG luminosities found by direct SED integration are around 0.5dex lower than previously suggested by \citet{clark2005massive}. In that work, \citeauthor{clark2005massive} used the correlation between OI 7774 and $M_{\rm V}$ found by \citet{ferro2003revised} for A--G type stars, and adopted a $BC_{\rm V}$ of 0. This resulted in luminosities for the YHGs of \logl$\sim$ 5.7, placing the stars above the Humphreys-Davidson limit for cool SGs, consistent with a very young cluster age. The discrepancy with our results can be largely explained by the bolometric correction used by Clark et al., which was 1mag lower than that typically found for F supergiants \citep{koornneef1983near}. This would cause Clark et al.\ to overestimate the luminosities of the YSGs by 0.4dex. In addition, as noted by \citeauthor{clark2005massive}, the OI line strengths of the Wd1 YSGs go beyond the range of the calibration by \citet{ferro2003revised}. A simple extrapolation of this trend would likely introduce a further systematic shift, though we are unable to say in which direction this would go.


\begin{table*}
\renewcommand{\arraystretch}{2.0}
\centering
\caption{The bolometric luminosities for the 4 RSGs in Westerlund 1. Spectral types listed are from \citet{clark2010serendipitous}. }
\begin{tabular}{lllccccc}
\hline\hline
Star &  RA ($\deg$) & Dec ($\deg$) & SpT &\multicolumn{2}{c}{log($L_{\rm bol}$ / \lsun)} \\ [0.5ex] 
& & & &  SED & BC$_{\rm K}$\\ [0.5ex]
\hline
W237&16 47 03.1 & -45 52 19.0&M3Ia & 5.31$^{+0.10}_{-0.10}$ & 5.35$^{+0.18}_{-0.14}$ \\
W20 &16 47 04.7 &-45 51 24.0 &M6I  & 4.96$^{+0.11}_{-0.10}$ & 5.17$^{+0.22}_{-0.19}$ \\
W75 &16 47 08.9 &-45 49 58.5 &M4Ia & 4.78$^{+0.15}_{-0.12}$ & 4.91$^{+0.18}_{-0.14}$ \\
W26 &16 47 05.4 &-45 50 36.6 & M5  & 5.44$^{+0.11}_{-0.08}$ & 5.46$^{+0.16}_{-0.12}$ \\
\hline
\end{tabular}
\label{tab:lbols}
\end{table*}

\begin{table*}
\renewcommand{\arraystretch}{2.0}
\centering
\caption{The bolometric luminosities for the 6 YSGs in Westerlund 1. Spectral types are from \citet{clark2019census}. }
\label{tab:lbols_ysgs_tab}
\begin{tabular}{lllcccc}
\hline\hline
Star &  RA ($\deg$) & Dec ($\deg$) & SpT &log($L_{\rm bol}$ / \lsun) \\ [0.5ex] 
\hline
W4 & 16 47 1.42&	-45 50 37.1 & F3Ia & 5.05$^{+0.13}_{-0.21}$ \\
W8a &16 47 4.79& -45 50 24.9& F8Ia & 4.74$^{+0.13}_{-0.18}$\\
W32  &16 47 3.67&-45 50 43.5 & F5 & 5.22$^{+0.14}_{-0.24}$\\
W16a  &16 47 2.21&-45 50 58.8 & A5 & 4.96$^{+0.15}_{-0.26}$\\
W12a  & 16 47 6.61 &-45 49 42.1& F1 & 4.89$^{+0.15}_{-0.20}$\\
W265  &16 47 6.21 &-45 49 23.7& F1 & 4.89$^{+0.19}_{-0.21}$\\
\hline
\end{tabular}
\end{table*}

%

\begin{figure*}
\centering  
\includegraphics[width=0.49\textwidth]{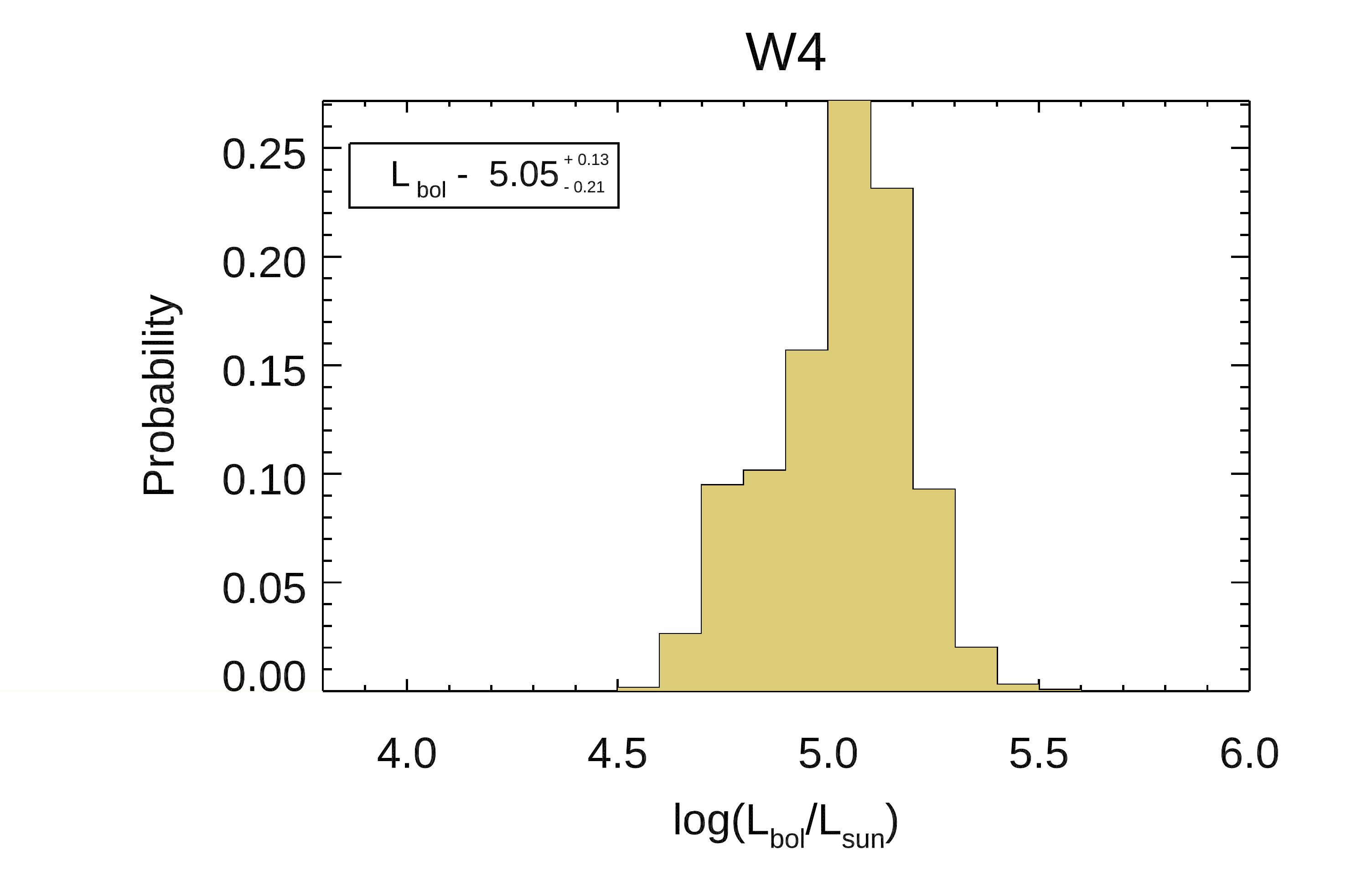}
\includegraphics[width=0.49\textwidth]{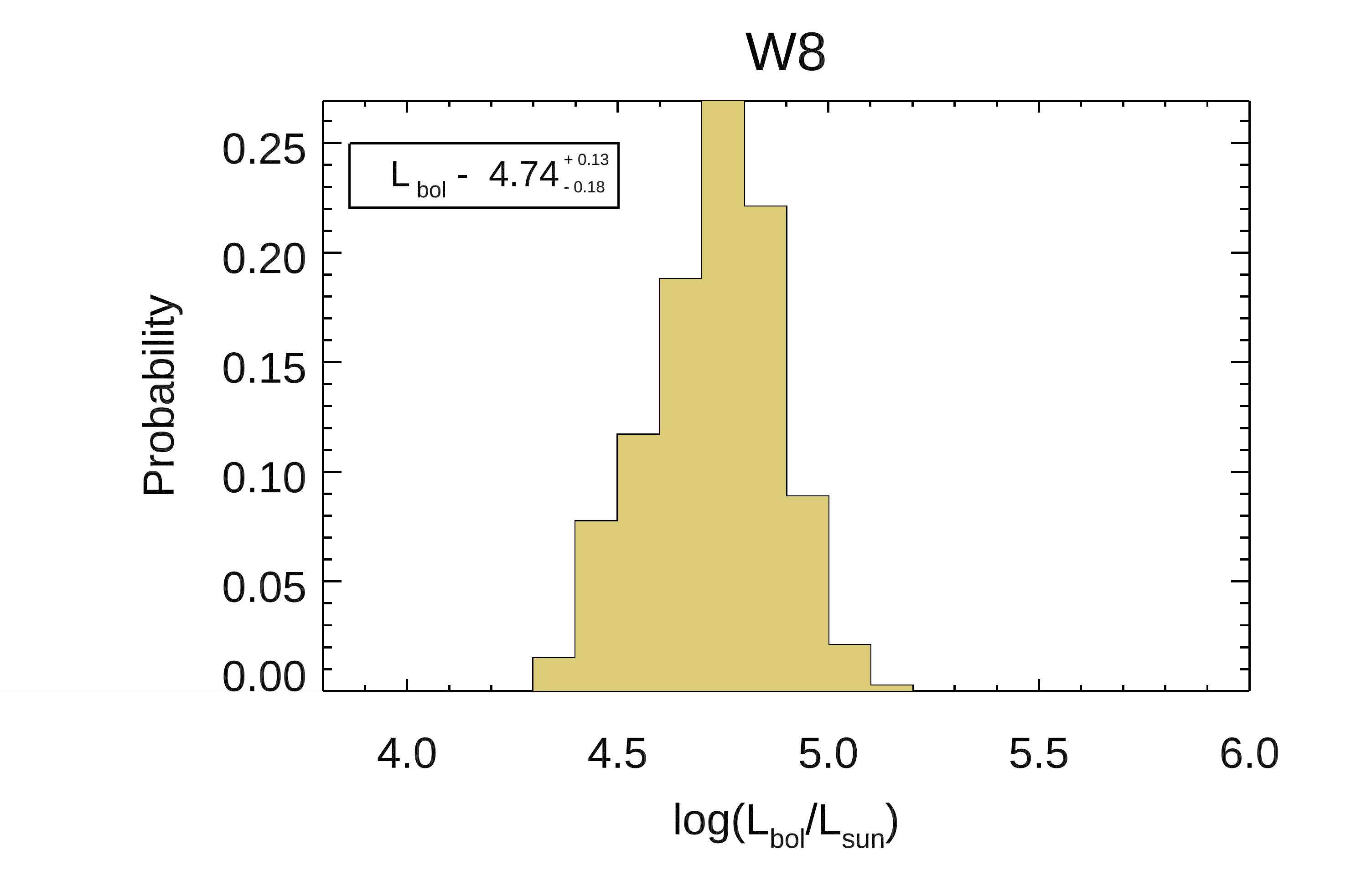}
\includegraphics[width=0.49\textwidth]{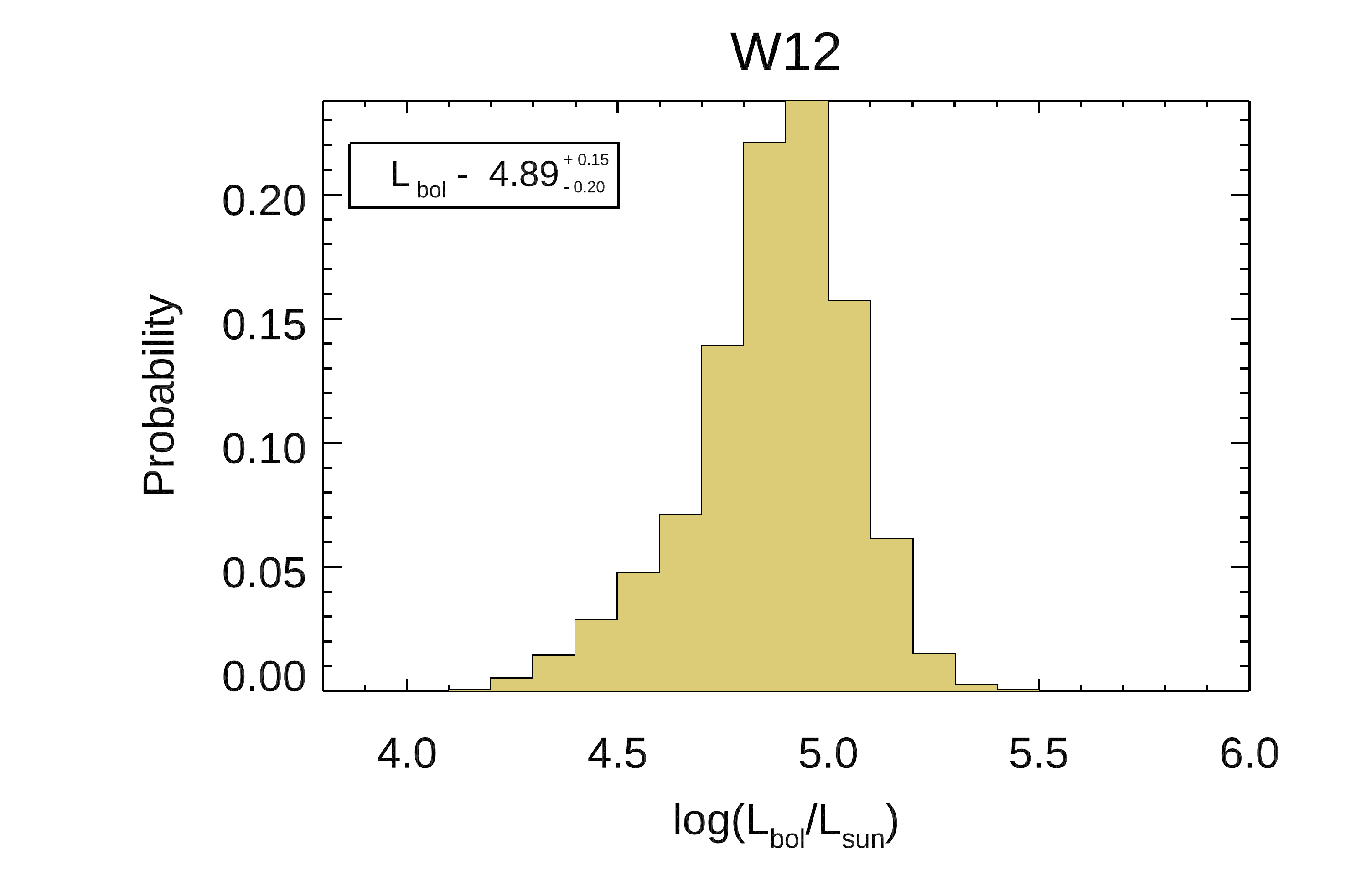}
\includegraphics[width=0.49\textwidth]{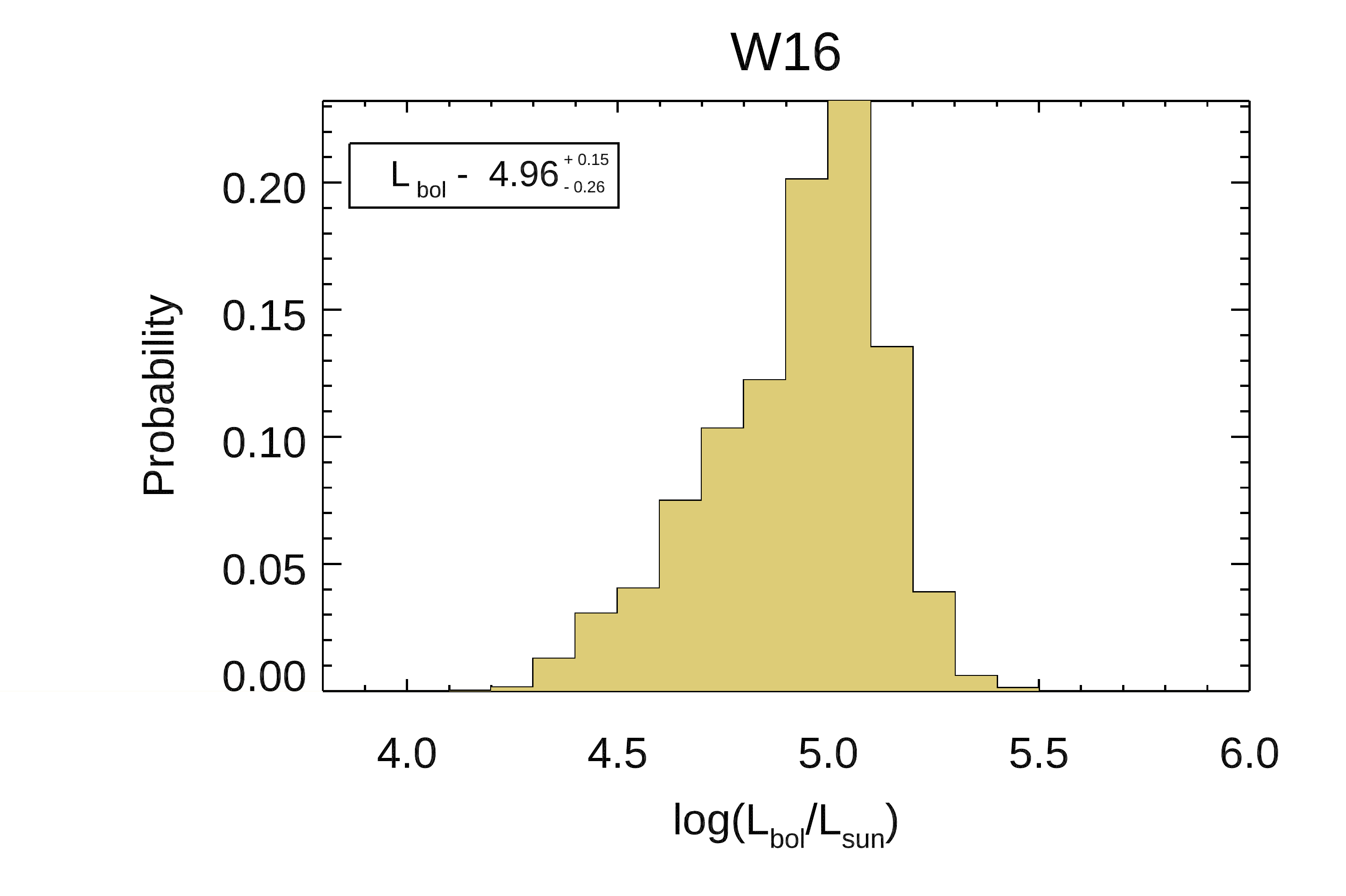}
\includegraphics[width=0.49\textwidth]{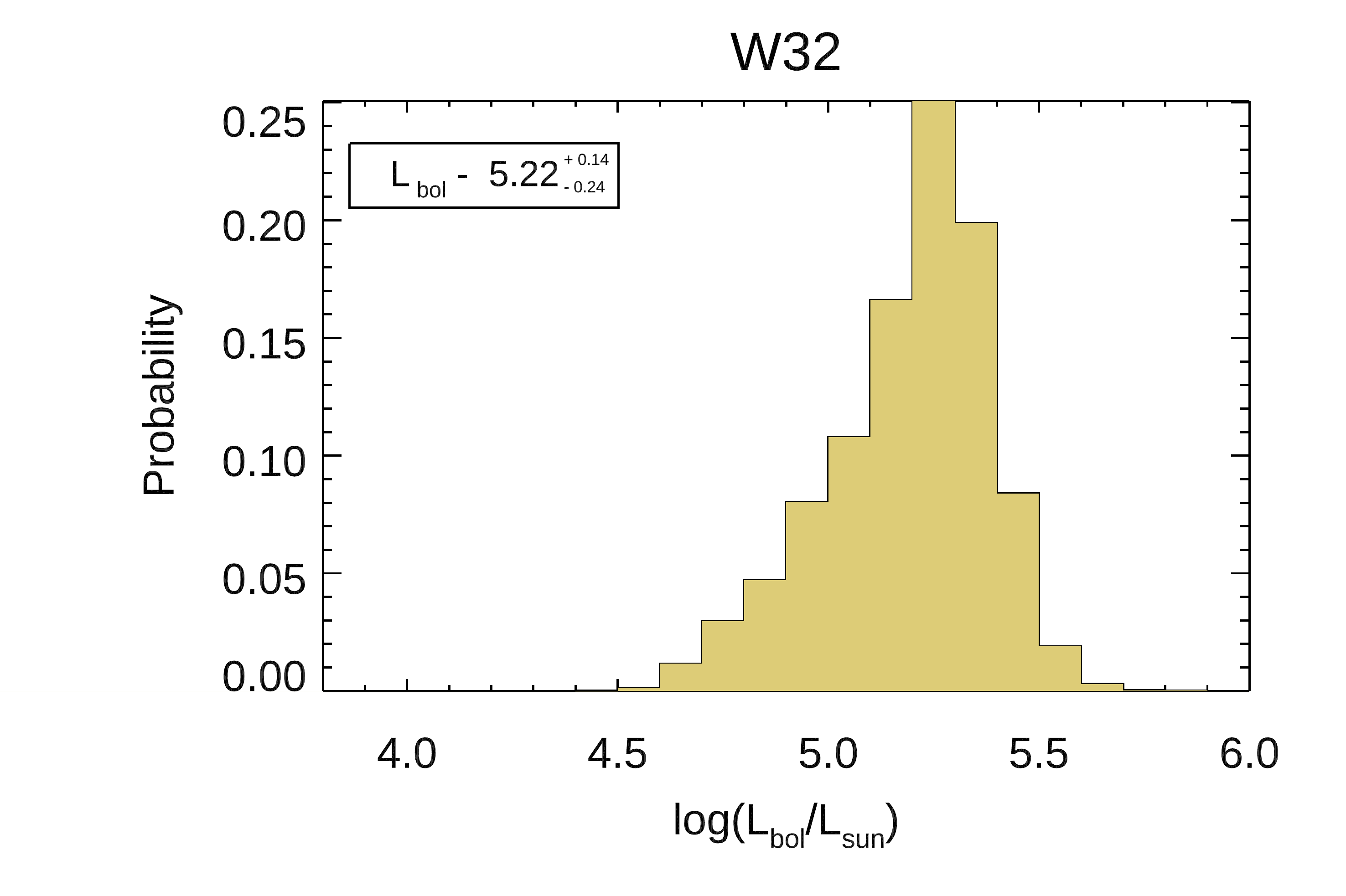}
\includegraphics[width=0.49\textwidth]{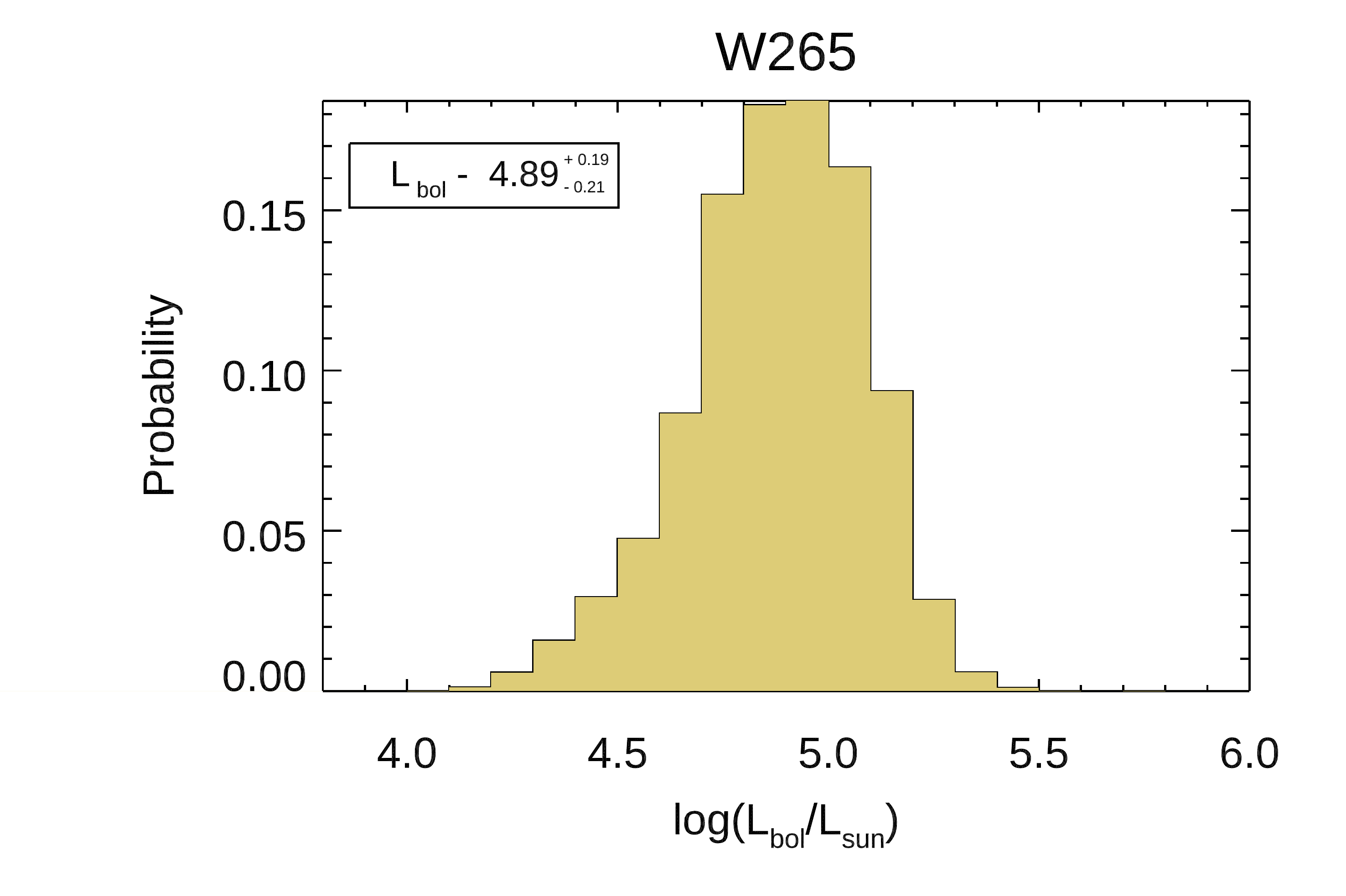}
\caption{The luminosity distributions obtained for each YSG in the cluster.}
    \label{fig:ysglbols_pdf}
\end{figure*}

\begin{figure*}
    \centering
    \includegraphics[width=15cm]{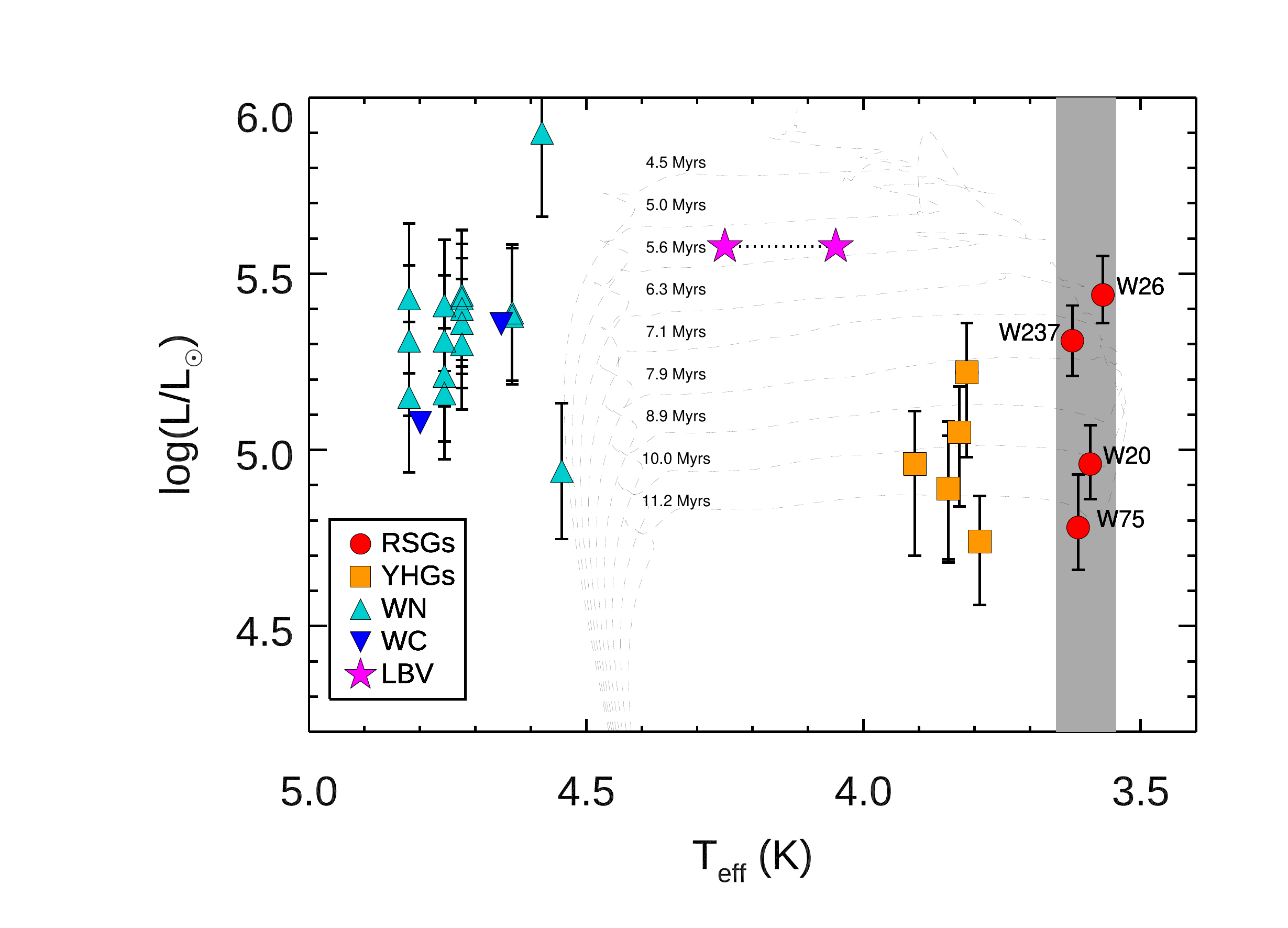}
    \caption{HR diagram for stars in Westerlund 1. The red circles represent the RSGs and the yellow squares represent the YSGs for which the luminosities have been determined in this work. As we do not know the effective temperatures of the stars they are offset from each other for clarity. The grey region shows the typical temperature range for RSGs. We also show data from \citet{aghakhankoo2019inferring}, whereby the luminosities of the hot stars \citep[originally from][]{crowther2006census,fenech2018alma} have been corrected for the distance used here. Overplotted are MIST isochrones \citep{dotter2016mesa,choi2016mist}.}
    \label{fig:HRdiagram2}
\end{figure*}

\section{The HR diagram of Wd1} \label{sec:hr}
In Figure \ref{fig:HRdiagram2} we present the updated HR diagram for the observed massive stars in Wd1. Also plotted in the diagram are MIST isochrones\footnote{MESA Isochrones \& Stellar Tracks, http://waps.cfa.harvard.edu/MIST/} for a range of ages \citep{choi2016mist}, which assume single star evolution and no rotation. It is immediately obvious that the cool supergiants (CSGs) in Wd1 are inconsistent with an age of 5Myr. The average luminosity of the CSGs is around $\log{L/L_\odot} = 5.1$, which contrasts with the theoretical expectation of $\log{L/L_\odot} = 5.6-5.7$. We note that all major single-star evolution codes predict the same CSG luminosity at this age to within $\pm0.1$dex. 

\begin{figure*}
    \centering
    \includegraphics[width=1.7\columnwidth]{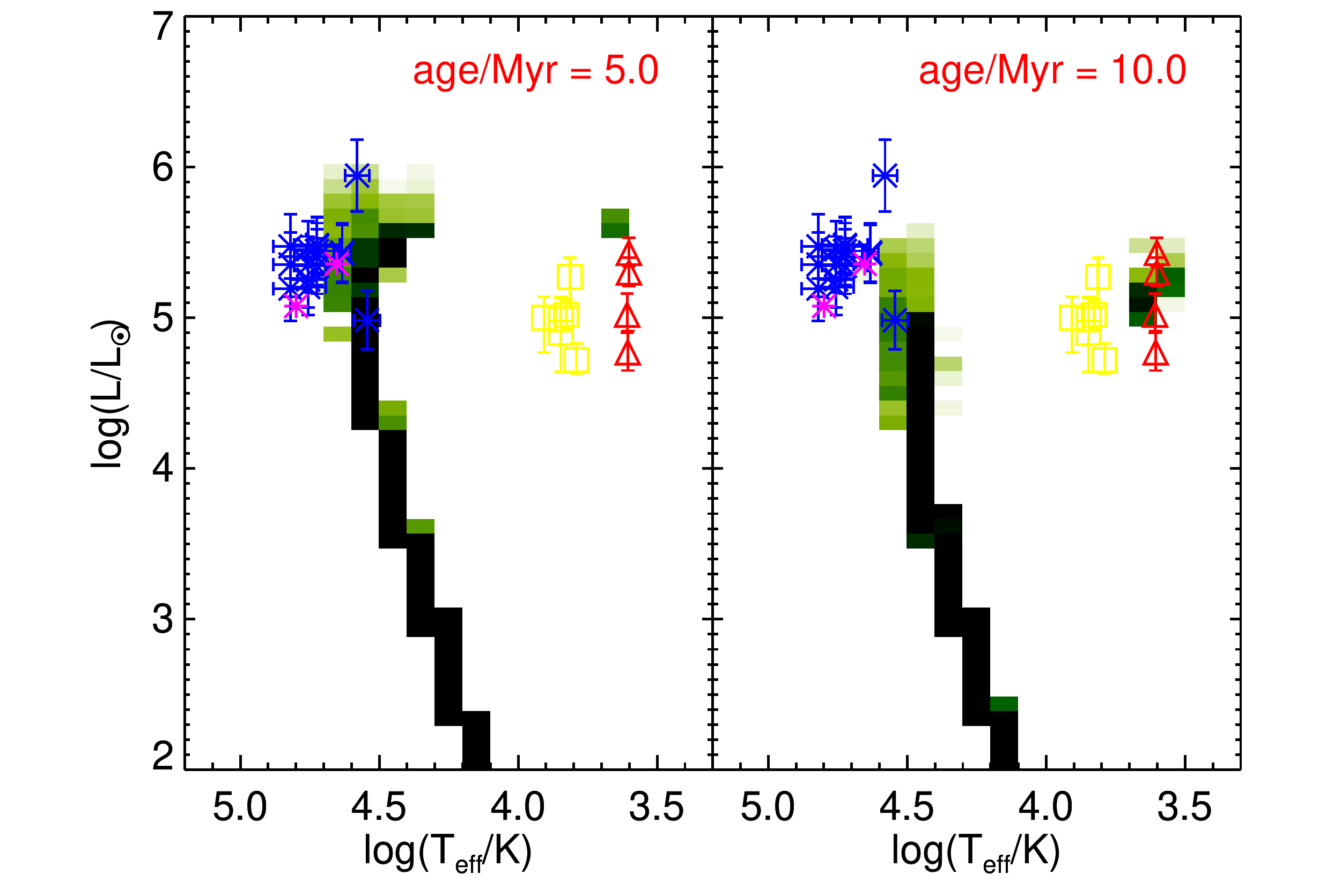}
    \caption{The HR diagram for Wd1, showing the locations of the RSGs (red triangles), YSGs (yellow squares), WNs (blue crosses), and WCs (magenta stars), in comparison to the predictions of BPASS binary population synthesis, for two ages: 5Myr (left panel) and 10Myr (right panel). The model predictions are plotted as 2-D histograms of stellar density, in arbitrary units.  }
    \label{fig:HR_bpass}
\end{figure*}

We can also compare the CSG luminosities to the predictions of binary star evolution. For this test, we use the Binary Population and Spectral Synthesis (BPASS) models \citep{stanway-eldridge2018}, the same as those used by \citet{dorn2018stellar} to infer an age of 5Myr from the number ratios of post main-sequence stars. In Fig.\ \ref{fig:HR_bpass} we plot BPASS H-R diagrams for two ages, 5Myr and 10Myr, and compare to the locations of the WRs and CSGs. As with single star models, BPASS binary models predict that at an age of 5Myr the CSGs should have luminosities of $\log(L/L_\odot) = 5.6\pm0.1$, well above what we observe. At 10Myr, the luminosities of the CSGs are much more consistent. We note that neither age isochrone, whether single star or binary, reproduces YSGs in the numbers observed.


\subsection{Formally determining cluster age from the cool supergiants}
From a simple comparison to predictions from evolutionary models of both single and binary stars, it is clear that the RSGs and YSGs are inconsistent with an age of 5Myr. Indeed, both suggest ages closer to 10Myr. In this section, we employ an adaptation of the `faintest RSG' method outlined in \citet[][ hereafter B19]{beasor2019ages} to derive a formal age with uncertainties. 

As discussed in B19, the various effects of binary evolution and/or stellar rotation may affect standard age determinations that employ model isochrones. Binarity and rotation may increase the brightness of the main-sequence turn-off, the brightnesses of the post-MS objects, as well as impact the relative number counts of objects in the various post-MS phases. B19 argued that, in a populous cluster containing numerous RSGs, the faintest RSG should be the most likely to have evolved in isolation, with a low rotation speed, and have arrived in the RSG phase most recently (and so be less affected by uncertainties in RSG mass-loss). Therefore, by comparing the luminosity of the faintest RSG to that expected from a population synthesis simulation employing single-star non-rotating models, one may obtain an age estimate which is least impacted by the various systematic effects which arise from the largest unknowns in massive stellar evolution (rotation, binarity, and mass-loss). 

More recently, \citet{eldridge2020ageing} used the BPASS models to test the use of RSGs as binary-independent age indicators, using both the mean RSG luminosity and the minimum RSG luminosity in a cluster to determine an age. This work argued that both the mean {\it and} the mininum RSG luminosity are roughly the same when using either binary or single star models (see Fig. 1 within), with slightly better agreement between input and output when the mean luminosity was employed. However, for this method to apply directly to real clusters, BPASS would have to accurately reproduce the luminosity evolution of stars in the RSG phase (i.e. how long each star spends at a given luminosity), something which all evolutionary models disagree on. The method described in B19 only requires models to be able to predict the luminosity of non-rotating single stars at the point they enter the RSG phase as a function of initial mass, a quantity which has relatively low variance from model to model. Nevertheless, here we use both methods to estimate the age of Wd1.  

For the current study, we make one adjustment to the B19 method. In Wd1, the lowest luminosity RSG is W75 at \lbol=4.77$^{+0.14}_{-0.12}$. However, the extinction map in \citet{damineli2016extinction} suggests that W75 sits in an area of high foreground extinction. Indeed, the SED of W75 appears reddened at 1$\micron$, as shown in Fig. \ref{fig:photometry}. This reddening is unlikely to be caused by circumstellar material as the lost flux does not appear to be re-emitted in the mid-IR. Due to the uncertainty around the true luminosity of W75, we adapt the 'faintest RSG' method of B19 to instead estimate the age from the 2nd faintest RSG. That is, we use population synthesis to estimate the most likely luminosity of the 2nd faintest RSG in a cluster of single non-rotating stars, and compare to the luminosity of the 2nd faintest star in Wd1 (W20). We note that an anomalously high extinction for W20, which would bias our analysis to an older age, can be ruled out by the star's colours. Specifically we place an upper limit on W20's extinction of $A_K=0.76$ based on its optical photometry under the assumption that its spectral type was later than M0 at the time of observation (see also Sect.\ \ref{section:rsglbols}).

\begin{figure}
    \centering
    \includegraphics[width=\columnwidth]{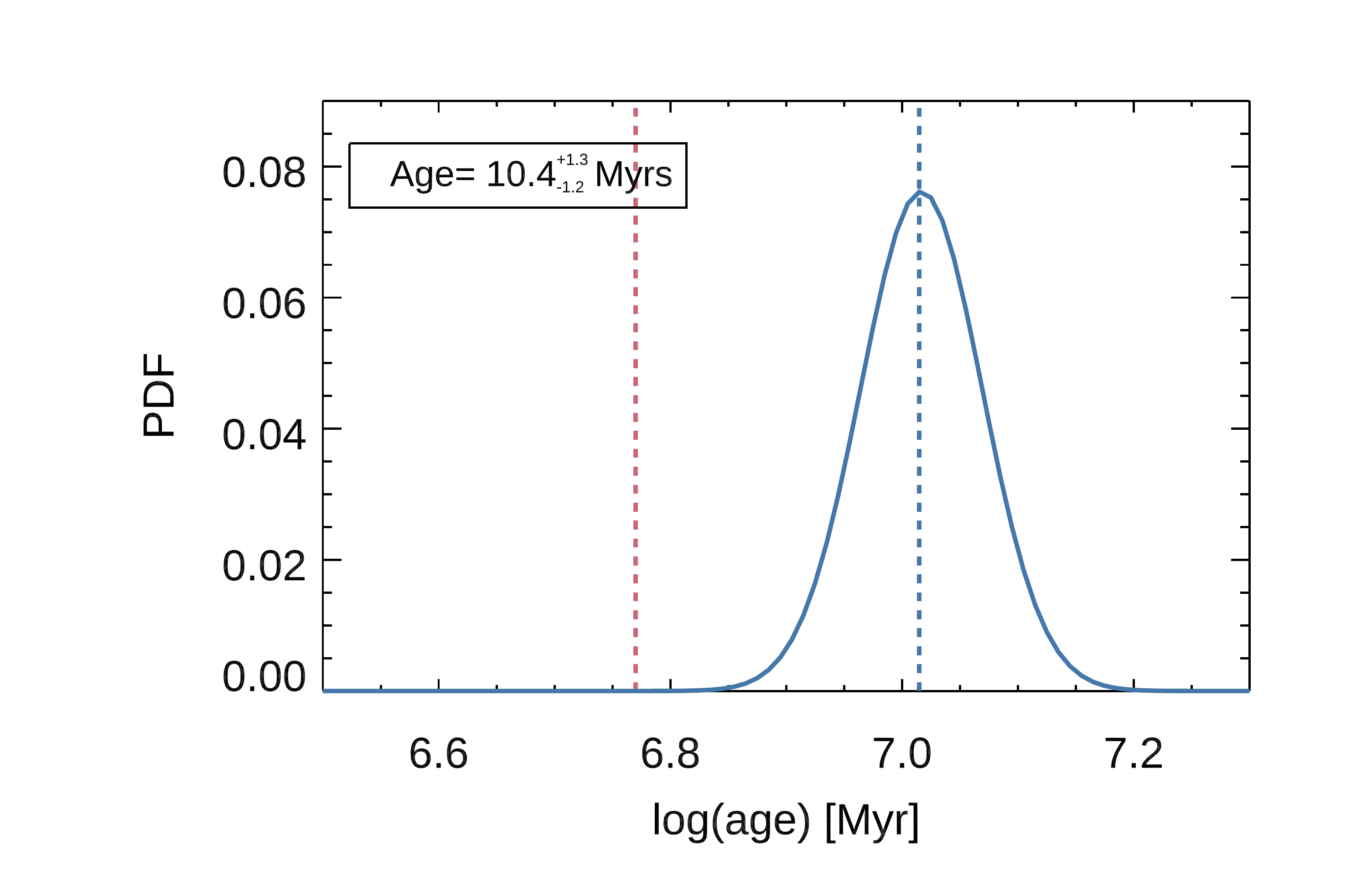}
    \caption{Probability distribution function (PDF) for the age of Wd1 (solid blue line). The blue dashed line shows the age found in this work employing the faintest RSG method, while the red dashed line shows an age of 6 Myr.}
    \label{fig:age_pdf}
\end{figure}

As the number of RSGs in a cluster is low (on the order of $\sim$10), the `2nd-faintest RSG' method is subject to stochastic errors. For this reason, we employ a probabilistic methodology in order to estimate the magnitude of these errors. We utilise the MIST grid of stellar isochrones \citep{choi2016mist} and for each age we isolate the RSG phase (defined as where $T_{\rm eff} \le$ 4500K and $\log (L_{\rm bol}/ L_\odot) \ge$ 4). From this we determine the minimum and maximum mass of star in the RSG phase and generate a sample of 500 synthetic stellar masses following a Salpeter IMF \citep{salpeter1955}. We then take a random subsample of 4 RSGs (to match the number of RSGs in Wd1) and take the second least luminous RSG, $L_{\rm 2min}$. This process is repeated 1000 times, from which we determine a relation of most likely $L_{\rm 2min}$ vs isochrone age. This is then compared to the \lbol\ of W20. To take into account the errors on W20's \lbol, we randomly sample from the luminosity distribution of that star (shown in Fig.\ \ref{fig:lbol_dists}) and repeat for 1000 trials, generating a PDF of possible ages shown in Fig.\ \ref{fig:age_pdf}. From the median of the PDF, we arrive at an age for Wd1 of \rsgage, corresponding to an initial mass for the RSGs of $17^{+3}_{-2}$\msun. We emphasise that this age estimate incorporates the uncertainties due to distance, extinction, as well as stochastic effects arising from a small sample size. 

For completeness, we also take the mean of the logarithmic luminosities of RSGs in Wd1, $\log(L / L_\odot)$ = 5.1, and use the BPASS relation of mean RSG luminosity as a function of age \citep[see Table 2 within][]{eldridge2020ageing}. We find an age of $11\pm1$Myr, consistent with what one would expect from visual inspection of Fig.\ \ref{fig:HR_bpass}, and within the errors of our estimate above.   

One possible source of systematic error in this method is the evolutionary state of W20. The B19 technique inherently assumes that the faintest (or in this case, the 2nd faintest) RSG will be relatively unevolved. Whilst W20 has no apparent maser emission \citep{Fok2012}, which is thought to be a signpost of an evolved evolutionary state \citep{davies2008cool}, it does have some mid-IR excess indicative of a sustained period of cool-wind mass-loss (see Fig.\ \ref{fig:photometry}). In principle, the non-zero likelihood of the fainter RSGs in a cluster being advanced in their evolution is reflected in the probability distribution function on the cluster age. However, if the RSGs in Wd1 were all closer to the end of the evolutionary track than would be expected on average, this would cause the age estimate described above to be systematically underestimated, making the cluster older than 10Myr. 

As was already clear from a simple qualitative comparison to model predictions, the age inferred from Wd1's cool supergiants is significantly older than previous estimates (discussed in Section \ref{sec:disco}), which tend to cluster around 5 Myr. Quantitatively, the B19 method excludes ages younger than 6Myr at the 99.9\% confidence level. If the faintest RSGs are more evolved than would be expected from a simple random sampling of a stellar isochrone, as indicated by their maser emission, this would make the disagreement with previous works even worse. Other possible sources of systematic error which may reduce this disagreement are discussed in the next section.

\subsection{Can systematic effects reconcile our observations with previous age estimates?}\label{section:clusterage}

\subsubsection{Variability}
RSGs are well-known to be variable, with the level of variability being wavelength dependent. In extremes, the variability in the $V$, $R$, $K$ and mid-IR seems to have an amplitude of around 1, 0.5, 0.25 and $<$0.1 mags respectively \citep{levesque2007late,yang2018red,ren2019period}. It is not clear whether this variablity is bolometric (i.e. synchronised across all bands), or whether it is a manifestation of atmospheric temperature changes at constant luminosity. 

In this section, we investigate the possibility that the RSGs in Wd1 are on average much brighter than our estimates here, due to the epochs of our photometry coinciding with the minimum state of their photometric variability. To do this, we perform a Monte-Carlo (MC) experiment in which we determine the probability that we have caught the RSGs in Wd1 in a minimum state. For the purposes of this experiment, we make the {\it extreme} assumption that RSGs vary bolometrically at a level of 1.0 mags. We also assume that the RSGs in Wd1 have a time-averaged luminosity consistent with the expectations of a 5Myr old cluster, which according to BPASS models (see Fig. \ref{fig:HR_bpass}) is $\log(L/L_\odot) = 5.6\pm0.1$. First, we randomly sample four luminosities from a Gaussian distribution centred on $\log(L/L_\odot) = 5.6$ with a standard deviation of 0.1dex. Next, we assume that each star varies in luminosity sinusoidally, with an amplitude of 0.5mags (i.e.\ a variation in luminosity of $dL_{\rm var} = 0.2$dex, or a difference of 0.4dex in luminosity between the minimum and maximum states), and randomly sample from that distribution. As in Section \ref{section:rsglbols}, we then randomly sample from the probability distributions for extinction (for each of the four stars) and for distance (applying the same randomly generated distance to all objects in that trial). Finally, we repeat $10^5$ times and record the posterior probability distribution of the average luminosity. 

\begin{figure}
    \centering
    \includegraphics[width=\columnwidth]{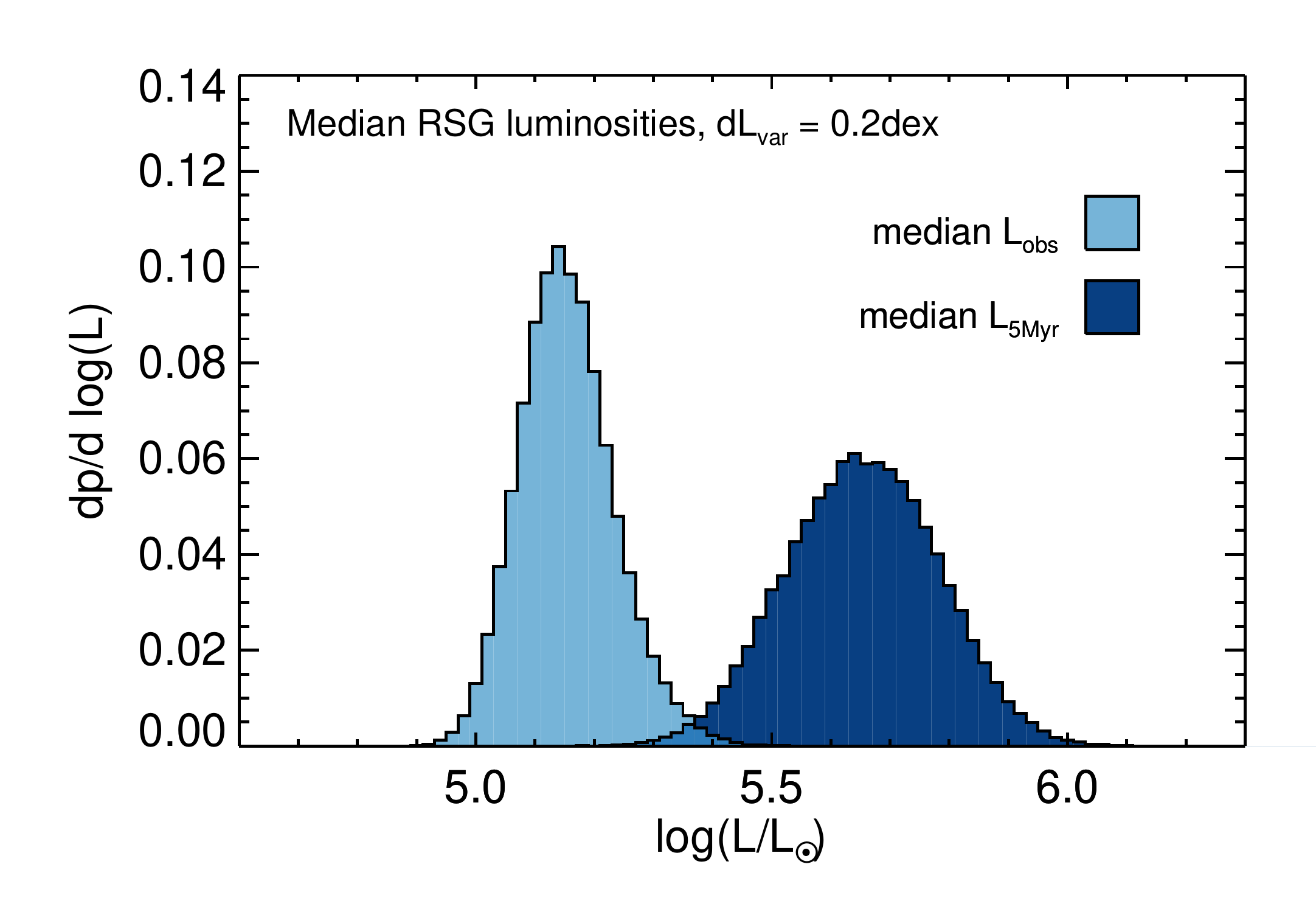}
\caption{The probability distributions of the median RSG luminosity from our observations (light blue) and from theoretical expectations if Wd1 were 5Myr old {\it and} all RSGs were bolometrically variable at the level of $\pm$0.5mag (dark blue).  }
    \label{fig:var}
\end{figure}

To compare to the results of our MC experiment, we generate a second probability distribution, this time for the median observed RSG luminosity. We take each star's probability distribution on luminosity (shown in Figs. \ref{fig:lbol_dists}-\ref{fig:ysglbols_pdf}), and perform a separate MC simulation where we sample from these distributions  $10^5$ times and determine the mean of the two middle luminosities (i.e. the median in the case of an even-numbered sample size).

The results of our MC experiment are shown in Fig.\ \ref{fig:var}. The peak of the observed distribution, and that of the expected distribution for a 5Myr old cluster, are obviously well separated. The width of the expected distribution is determined mostly by the level of variability of the RSGs. In the extreme case discussed here (bolometric variability of $\pm$0.5 mag, or $\pm$0.2dex in luminosity), the probability that we have caught all RSGs in a low state such that the average luminosity is $\sim$0.4dex below the median state can be calculated from the overlapping region of the two histograms. Specifically, we find this probability to be 2\%.

Whilst low, the above-derived probability is not conclusive insofar as it is not outside 3$\sigma$. However, we reiterate that our experiment employed an extreme assumption for RSG variability, and that {\it all} RSGs in Wd1 were variable at this level. Furthermore, since the photometry we used to define each star's SED was collected non-contemporaneously, we have effectively made another extreme assumption that each RSG was caught in a low state in every photometric epoch. These factors considered, we consider the $2\%$ probability we estimate here to be an extreme upper limit, and so conclude that stellar variability {\it cannot} explain the disagreement between the luminosities of Wd1's RSGs and the expectations for a 5Myr old cluster. 


\subsubsection{Extinction law} \label{sec:hr_extinct}
Throughout this work we have utilised the extinction law determined by \citet{damineli2016extinction}, which covers wavelengths between 0.4 -- 8.0$\mu$m. However, there are a number of different extinction laws presented in the literature. To assess the systematic impact our choice of extinction law choice on the luminosities of the CSGs, we repeat the above analysis using relations from \citet{cardelli1988ext}, \citet{nishiyama2009interstellar} and \citet{hosek2018optical}, still assuming an $A_{\rm K}$ value of 0.74$\pm$0.08 \citep{damineli2016extinction}. The \citeauthor{nishiyama2009interstellar} and \citet{hosek2018optical} laws cover spectral ranges 1.2 -- 8 $\mu$m and 0.8 -- 2.2 $\mu$m respectively, and so for shorter wavelengths we splice the relations with that of \citet{cardelli1988ext}. In doing this, we find the effect of different extinction laws causes only a modest change in luminosity, by a maximum of 0.1 dex. The star for which the different extinction laws have the largest systematic effects on \lbol, W20, has a maximum luminosity of \logl = 5.02 when using the \citeauthor{damineli2016extinction} law, and a minimum luminosity of \logl = 4.93 when using the \citeauthor{cardelli1988ext} law. Indeed, the \citet{damineli2016extinction} relation has the steepest $A_{\rm V}/A_{\rm K}$, and thus results in the highest \lbol\ values out of all the extinction discriptions investigated. We therefore conclude that varying the extinction law cannot raise the CSG luminosities enough to allow them to be consistent with an age of 5 Myr.


Interestingly, the most recently derived extinction law in the sightline of Wd1 by \citet{hosek2018optical} also suggests an age $>$5Myr for this cluster may be likely. These authors note that the precise nature of their extinction law depends upon the assumed age and distance for Wd1. To break this degeneracy, the \citeauthor{hosek2018optical} use the distance estimated from the eclipsing binary system W13 \citep[3.9 $\pm$ 0.4 kpc][, consistent with the distance we use in this study]{koumpia2012fundamental}. Ultimately, Hosek et al.\ find this result favours an age for Wd1 $\ga$7Myr (see Fig. 16 within).

\section{A critical look at previous age estimates}\label{sec:disco}
In the previous section, we have shown that the luminosities of the cool supergiants in Wd1 are inconsistent with a cluster age of 5Myr. Furthermore, we have argued that this tension cannot be solved by systematics in our analysis. In this section, we take a detailed look at the other age estimates for Wd1 in the literature, to see if they can be reconciled with our own. 

\subsection{Stellar diversity}
We first consider estimates which use the variety of massive stars present in the cluster as an age constraint. \citet{clark2005massive} suggest an age between 3.5 - 5 Myr, constrained by the presence of both WR stars and O stars. Under the paradigm of single star evolution, Clark et al.\ argued the presence of WCL stars puts a lower age limit on the cluster of 3.5 Myr, while the O supergiants suggest a maximum age of 5 Myr \citep{clark2005massive,meynet2003stellar}. \citet{crowther2006census} detailed a census of the WR population of Wd1 using near-IR photometry and spectroscopy. Using the observed ratio of WR stars to RSGs and YSGs, they favour an age of 4.5-5 Myr, though they note that neither single-star nor binary evolution models are able to reproduce the WN/WC ratio at any age. More recently, \citet{dorn2020populations} use Wd1's number ratios various stellar classes (e.g. WR/RSG) and compare to predictions from BPASS, finding a most-likely age of 5Myr. 

As we have already shown in Figs.\ \ref{fig:HRdiagram2} and \ref{fig:HR_bpass}, 5Myr isochrones substantially overestimate the luminosities of CSGs in Wd1. Indeed, both single star and binary evolution models require that the age of Wd1 is closer to 10Myr to be able to fit the CSG luminosities. This leads us to conclude that single-age evolutionary models {are incapable} of simultaneously reproducing Wd1's relative number counts of post-MS objects {\it and} the luminosities of the YSGs and RSGs. 

In general, any method which employs the ratios of star counts in different evolutionary phases as the age diagnostic is fundamentally limited by how well the adopted evolutionary model can reproduce the evolution of stars through those phases. For example, in the case of \citet{dorn2020populations}, the accuracy of their inferred age depends on the ability of BPASS to model stars' luminosities, effective temperatures and surface abundances as a function of time. These factors, which are extremely sensitive to binarity, rotation, and mass-loss, ultimately govern which stars become WRs and RSGs \citep[as defined by BPASS, see][]{dorn2020populations} and for how long. This in turn determines the ratio of WRs to RSGs as a function of time. This ultimately means that it will be extremely challenging for any stellar evolution model to use number ratio of WR/RSG as a precise age diagnostic. The strength of the B19 method is that it only requires the evolutionary model to predict the minimum expected luminosity of a cluster's RSGs (which corresponds to that of the main-sequence turn-off) as a function of time for non-rotating single stars, which most evolutionary codes seem to agree on.

\subsection{Isochrone fitting}
The time a star takes to contract to the main-sequence is a strong function of its mass. By estimating the brightness of the pre-MS -- MS transition region, the luminosity (and hence mass) of star to have recently arrived on the MS can be deduced. The pre-MS contraction time of that star is then assumed to be the age of the cluster. Therefore, isochrone fitting to this region of the cluster's colour-magnitude diagram (CMD) can be used to estimate the age. 


The first such work on Wd1 was done by \citet{brandner2008intermediate}, which was later updated in \citet{gennaro2011mass}. In the latter paper, the authors argued for an age of 4$\pm$0.5 Myr. However, the authors do not perform a formal fitting procedure to determine the best-fit age, instead they identify by eye the most likely combination of distance and age to match the observed CMD, see Fig. 5 within. The error bar they quote on the age is half the spacing of their isochrone grid, rather than a formal description of the probability distribution. The same group studied Wd1 again in \citet{andersen2017very}, though in that work no formal age estimate was provided. 

A similar study was presented in \citet[][ hereafter K12]{kudryavtseva2012instant}. Here, the authors used high-precision proper-motion measurements to identify high-probability cluster members, in order to minimise the considerable errors that can be introduced by field star contamination (which we will discuss again later). These authors found an extremely narrow peak in the probability distribution at 5Myr. This led them to argue for an upper limit to any age spread of 0.4Myr. As we will argue below, this small uncertainty on age is misleading as there are several potential sources of systematic error which will likely dominate the formal fitting errors. 

\begin{figure}
    \centering
    \includegraphics[width=\columnwidth]{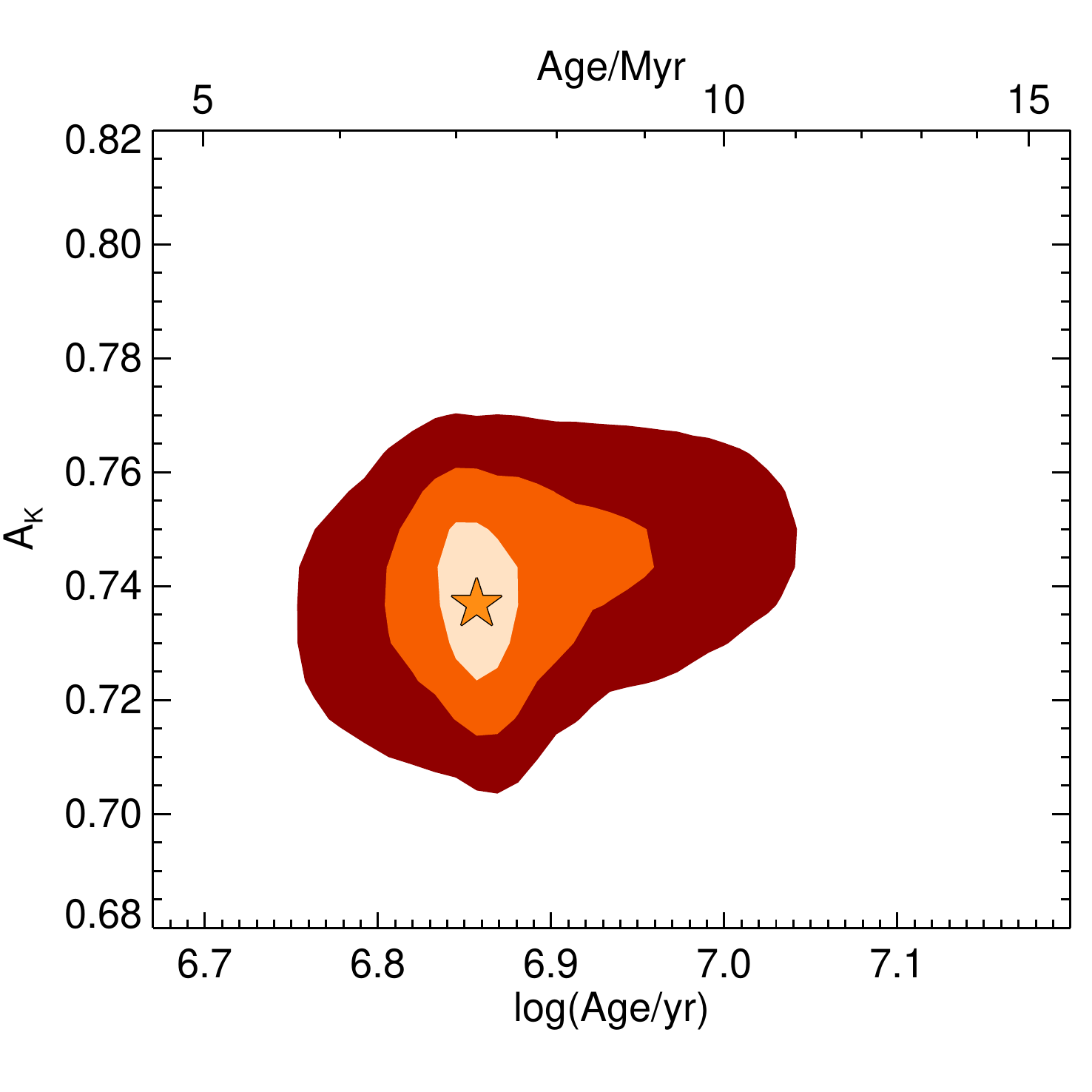}
    \includegraphics[width=\columnwidth]{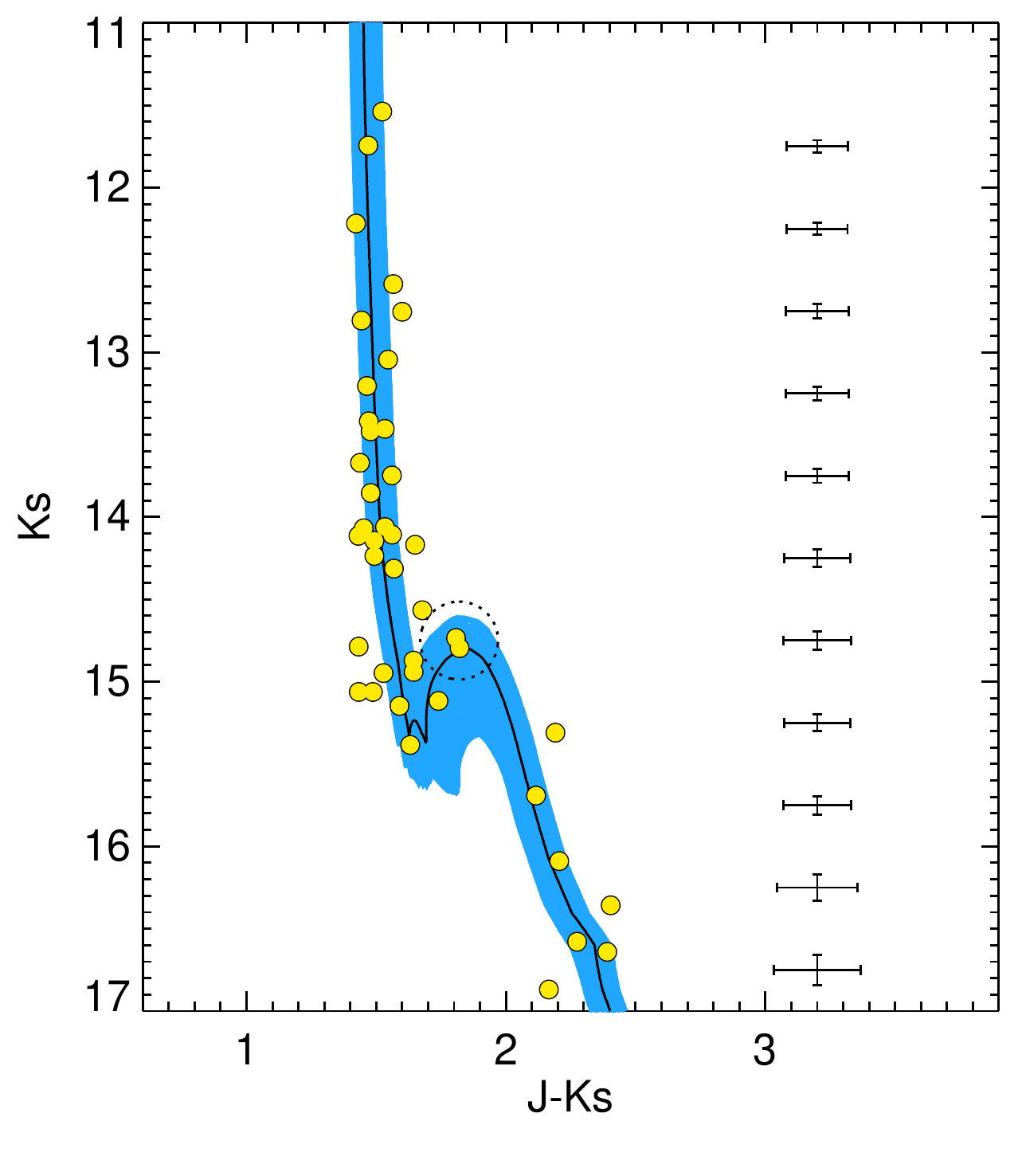}
    \caption{Results of fitting the K12 pre-MS photometry. {\it Top}: the $\chi^2$ solution plane of our analysis. The shaded contours indicate the 1, 2 and 3$\sigma$ confidence regions, while the star symbol indicates the best-fit solution. Note that the probability distribution does not account for the uncertainty in distance, which we treat as a systematic error. {\it Bottom}: The K12 photometric points (yellow), overlaid on the model isochrones which fall within the 95\% confidence limits. The errors on the photometry, which include that on extinction, are shown on the right of the panel. The dotted circle identifies the `bridge' stars discussed in the text.}
    \label{fig:K12}
\end{figure}

\subsubsection{A re-analysis of Wd1's pre-MS stars}
Here, we re-investigate the pre-MS age of Wd1. We begin with the dataset of K12, which is an astrometrically-cleaned sample of stars from Wd1. We fit the CMD with isochrones from \citet{Haemmerle19}, interpolated onto an age resolution of 0.02dex, though we find no systematic differences when the FRANEC isochrones employed by K12 are adopted. We use the extinction law of \citet{damineli2016extinction} to determine the reddening per unit extinction in the $J-K / K$ plane, and the distance derived in Sect.\ \ref{section:wd1properties}, both of which are slightly different to those employed by K12 (see Sect.\ \ref{sec:differences}).

One further difference between K12 and our analysis is how the errors on each photometric point are treated. In K12, the errors are assumed to be the formal photometric errors (hereafter the {\it instrumental} errors), which depend on the photon shot-noise and/or the background variance. In high quality data, such as that presented in K12, these errors can be very small compared to the distance in CMD-space of any point from the best-fitting isochrone (see left panel of Fig.\ 2 in K12). This leads to anomalously low un-normalised likelihoods for the best fitting model, and large differences between the likelihood of the best-fit model those of the neighbouring models. This in turn leads to very small formal errors, as the best-fit model has a likelihood which is overwhelmingly better than any other model in the grid. 

In reality, it is very unlikely that the instrumental errors are the dominant source of uncertainty on any data-point. This can be seen at the bright end of the CMD in K12's Fig. 2, where the dispersion of the photometry around the isochrone is far in excess of what would be expected from the photometric errors (indicated on the left-side of that plot). The probable cause of this is the presence of differential extinction, which is unlikely to be correlated with object brightness. 

From a simple calculation of the standard deviation of the brightest stars in the K12 sample, we estimate that the error due to differential extinction on colour is $\pm0.115$mag. To determine the error on $A_K$, we assume that the error on colour is the quadrature sum of those on $A_K$ and $A_J$, with $A_J = 3.23 A_K$ \citep[following ][]{damineli2016extinction}. This gives an error on $A_K$ due to differential extinction of 0.034mag. For each data-point, we then assume that the total error is the quadrature sum of the instrumental error and that from differential extinction. 

To find the best fitting model isochrone, we perform a free two-dimensional fit on age and extinction. For each isochrone age and extinction pair in our grid, we determine the shortest linear distance in CMD-space of each photometric point $i$ from the model isochrone in units of $\sigma$, where $sigma$ is the quadrature sum of each photometric point's error on magnitude and colour. For each age and extinction, we then determine the quantity $\chi^2 = \sum_i \sigma_i^2$, and identify the best-fitting age-extinction combination to be that with the lowest $\chi^2$. We determine the 68\% and 95\% confidence limits (analagous to 1-2$\sigma$ error bars) from the points in the grid with $\chi^2 < \chi^2_{\rm min} + \Delta\chi^2$, where $\Delta\chi^2 = 2.30$ and 6.18 respectively \citep{anvi1976energy}. Finally, we adopt a distance of 4.12kpc, and treat the uncertainty on distance as an external systematic error.

The results of our analysis can be seen in the panels of Fig.\ \ref{fig:K12}. Our best-fit {\it mean} extinction $A_K = 0.740 \pm 0.015$ is consistent with the $A_K = 0.74 \pm 0.08$ found by \citet{damineli2016extinction}. We note that our error is the error in the mean; we have already assumed an intrinsic dispersion on the mean $A_K$ of 0.034mag. Neglecting the error on distance, we find a best-fitting age of $7.2\pm0.2$Myr, which we plot in CMD-space in the bottom panel of Fig.\ \ref{fig:K12}. While the 1$\sigma$ errors are quite small, we note that the probability distribution is somewhat flared (see Fig.\ \ref{fig:K12}, bottom panel), with the 95\% confidence intervals extending from 6-9Myr. 

We emphasise that the results above do not include any uncertainty on distance. To estimate the systematic error due to distance, we simply repeat the analysis using distances at the 68\% confidence limits of 3.79kpc and 4.78kpc. In doing so, we find an additional distance-based systematic error on our best-fit age of $^{+1.1}_{-2.3}$Myr. The effect of other sources of error are discussed in Sect.\ \ref{sec:systematic}.

\subsubsection{Differences from Kudryavtseva et al.} \label{sec:differences}
Our best-fit age differs from that of K12 for several reasons. Firstly, the extinction law we use \citep[][, $A_J/A_K = 3.23$ ]{damineli2016extinction} differs substantially from that used by K12 \citep[][, $A_J/A_K = 2.52$]{rieke1985interstellar}. The extinction law determines the angle of the vector along which a model isochrone will be displaced across the CMD -- the steeper extinction law of \citet{damineli2016extinction} results in a smaller downward shift of the isochrone in CMD space at fixed colour shift, the latter being well-constrained from the position of the MS. This then results in older isochrones, which have an intrinsically fainter pre-MS/MS transition point, being required to fit the data. If we repeat our analysis using the Rieke \& Lebofsky extinction law, we find an age of $5.2_{-0.6}^{+0.2}$Myr, consistent with K12's measurement of $5.0\pm0.2$Myr. 


\subsubsection{Sensitivity of the pre-MS age to systematic errors} \label{sec:systematic}
The above analysis leads to an age estimate which seems overly-precise, with a formal error on the age of just $\pm0.2$Myr when the uncertainty in distance is ignored. However, implicit in this analysis is the assumption that every data-point in the CMD is randomly scattered away from the best-fitting isochrone according to its experimental errors alone. In reality there are a host of systematic errors which could affect any data-point, for example: unresolved binarity (which would make a star appear abnormally bright); anomalously high/low extinction (which would make the object too red/blue {\it and} to faint/bright); or imperfect foreground decontamination (which would remove the data-point from the CMD all together). 

Furthermore, certain stars in the CMD have substantially greater leverage over the best-fit solution than others. The brighter stars in the K12 sample with $K<14.5$ have almost zero diagnostic power over the inferred age, as the pre-MS is always fainter than this for all but very young ages ($\la$3Myr). Similarly, the fainter, redder stars ($K \ga 15.5$) are consistent with all ages younger than $\sim$20Myr. In fact, almost all leverage over the cluster age in K12's data comes from two stars just above the peak of the pre-MS region at $(J-K,K) \simeq (1.8,14.8)$ (hereafter the `bridge' stars, denoted by the dotted circle in the bottom panel of Fig.\ \ref{fig:K12}). In this Section, we explore the effect of systematic errors on the best-fit age and its uncertainty, with particular emphasis on the `bridge' stars. 

\paragraph{Anomalous extinction}
In our analysis, we have assumed that the dispersion of the photometry about the best-fitting isochrone is dominated by inhomogeneous extinction, which we quantify by measuring the standard deviation of the colours of the bright stars. We have then applied this uniformly to all stars, though in reality it is unlikely to affect all stars the same. If instead we allow the error on extinction to be larger for just the `bridge' stars, such that for these two stars $dA_k=0.06$mags \citep[cf.\ the error on the average cluster in $A_K$][]{damineli2016extinction}, the leverage these stars will have over the best-fit solution will be reduced for the following reason: when the errors on these points are small, their location on the CMD can only be explained by a bright pre-MS; but, when the errors are larger, they could just as easily be MS stars randomly scattered to a redder colour. The consequences of this can be seen when we repeat our analysis with the larger errors for the `bridge' stars. The best-fit age becomes $7.4^{+1.6}_{-0.4}$Myr; slightly older, but with a much broader probability distribution since the constraining power of the `bridge' stars is reduced.

\paragraph{Anomalous foreground contamination}
Next, we investigate the effect of treating the `bridge' stars as foreground contaminants. Removing one star from the analysis, the best-fit age remains the same, but the errors increase as age constraint is loosened, with older ages becoming possible ($7.2^{+1.4}_{-0.2}$Myr). Removing both stars results in an age of $7.6^{+1.8}_{-0.6}$Myr; marginally older, but with larger errors.

\paragraph{Unresolved binarity}
Finally, we repeat the analysis but with the assumption that just one of the `bridge' stars is an unresolved equal-mass binary, and replace the star with two of similar colour but half the brightness. Now, the probability distribution is pulled dramatically to older ages, the best-fit solution becoming $9.0^{+1.2}_{-0.6}$Myr.

\subsubsection{Pre-MS fitting: summary}
In our re-analysis of the K12 data, we have found that updating the extinction law to that measured by \citet{damineli2016extinction} moves the best-fit age from $\sim$5Myr to $\sim$7Myr. The formal 68\% confidence limits ($^{+1.1}_{-2.3}$Myr) are dominated by the uncertainty in the distance to Wd1. However, despite the K12 data containing over 40 putative cluster members, we caution that this best-fit age is strongly leveraged by just two data-points in the MS -- pre-MS transition region. Including extra sources of small systematic error to these two data-points can have a profound impact on the best fit age, and in all cases makes the age older and more imprecise.

As a final note on this topic, in our discussions on systematic errors in pre-MS fitting we have not mentioned the absolute level of accuracy that is possible from this type of analysis. The ages derived from the pre-MS ($\sim$7Myr) and the luminosity of the CSGs ($\sim$10Myr) are completely independent of one another in terms of the relevant stellar physics. For these measurements to be in tension with one another, each would require an absolute level of accuracy and precision of 10-15\%. In the case of pre-MS stars, it has been shown from nearby young clusters and associations that discrepancies in the theoretical mass-radius relation and/or synthetic photometry make this level precision extremely challenging \citep[e.g.][]{Naylor2009prems,David2019uppersco} 

Consequently, we conclude that while Wd1's pre-MS population indicates an age of 7Myr, ages older than 10Myr cannot be definitively excluded, and so any tension between the pre-MS and CSG ages is weak.

\subsection{Eclipsing binaries}
There are a handful of known eclipsing binary systems in Wd1, from which it is possible to estimate an age. This method requires measurements of the radial velocities (RV) and photometric light curves, from which dynamical masses can be determined provided an inclination ($i$) is known. An upper limit to the cluster's age is then determined from the expected lifetime of the most massive component in the system. This method has the advantage of being independent of distance, though it is extremely sensitive to the inclination $i$ of the binary system.

Of the four massive eclipsing binaries in Wd1, of particular interest are WR77o and W13, the two most massive systems. WR77o is a single-lined eclipsing binary where only one component, a WR, can be directly observed. For the invisible component, \citet{koumpia2012fundamental} determine a mass of $\sim$40\msun\ and propose this is a post-mass transfer system. However, we note that \citeauthor{koumpia2012fundamental} advise that their mass estimates for the components of WR77o should be taken with caution given the significant residuals in their best fits to the light curves, such that the authors themselves authors chose not to speculate further on the nature of the system.


A second high mass eclipsing binary, W13, has been studied in both \citet{ritchie2010vlt} and  \citet{koumpia2012fundamental}, with both authors finding similar mass estimates for the two components (23\msun\ and 33\msun). Ritchie et al.\ argue that the system is post mass-transfer, meaning that the initial mass of the more massive component was even greater than it is now. This would place a lower limit on Wd1's age of $\sim$5Myr, roughly the lifetime of a 33\msun\ star. 

Looking again at the data and its modelling in \citet{ritchie2010vlt}, there are two possible sources of systematic error. Firstly, Ritchie et al.\ measure the radial velocity variations in W13 by simple Gaussian fitting to spectral lines, rather than spectral deconvolution of the two components. Since the temperatures of the two stars in the system appear to be similar, one with lines in emission and one with lines in absorption, the two sets of spectral lines may cancel out, causing the radial velocity variations to appear smaller than they are. Secondly, Ritchie et al.\ obtain the key parameter of the system inclination by fitting the eclipse seen in the lightcurve of \citet{bonanos2007}. As noted by Ritchie et al., the eclipse is very weak (0.15mags), and is comparable to the intrinsic variability of stars similar to those in W13, which the authors do not factor in to their analysis. If included as an extra source of Gaussian noise on each photometric point in the light curve, it is likely that this would increase the error on the inclination, but also shift the best-fitting inclination to a lower value. This would systematically shift the masses of both stars in W13 higher, whilst also increasing the uncertainty. Therefore, It is unlikely that these effects would bring the W13 masses into better agreement with the cool supergiants.


\subsection{A summary of Wd1 age estimates, and the nature of Wd1}
To summarise our discussion in this section, we have studied all previous age estimates of Wd1 in detail, focusing on their systematic errors, to attempt to resolve the tension with the cool supergiant age of $10.4^{+1.3}_{-1.2}$Myr presented in this paper. Our conclusions are:

\begin{itemize}
    \item Age estimates from stellar diversity are unreliable. No evolutionary model (single star or binary, rotating or non-rotating) is able to simultaneously explain the presence of various post-MS objects {\it and} reproduce the luminosities for the cool supergiants. 
    \item We have re-analysed the pre-MS photometry of Wd1 from K12, finding an age slightly older than that inferred by those authors, 7.2$^{+1.1}_{-2.3}$Myr. The primary cause of this shift is our adoption of the Damineli extinction law. The formal errors are dominated by the uncertainty on distance. However, we find that almost all the leverage on age comes from two data-points close to the pre-MS -- MS transition region. We have shown that systematic errors on these two stars could substantially flare the probability distribution to the extent that ages above 10Myr move well within the 68\% confidence limits.
    \item Upper limits on Wd1's age inferred from the lifetimes of stars in eclipsing binaries are also likely subject to systematic errors. However, the systematics we discuss here are all likely to push the mass estimates higher, and hence pull the `eclipsing-binary' age lower. Specifically, the most massive component of the eclipsing binary W13 seems to place a hard upper limit on Wd1's age of 5Myr. This of course assumes that the more massive star is the mass-loser in the W13 system, which we do not discuss here. 
\end{itemize}

So, in terms of age estimates, the only substantial tension remaining is that between the brightnesses of the cool supergiants and the masses of the stars in W13. The presence of these objects in the same cluster cannot be explained by stellar evolution theory under the assumption that Wd1 is a single-age star cluster. 

Before going further, it is worth discussing Wd1 in context with other massive young star clusters. In particular, there is one other cluster in the Milky Way that has a similar mass and age to Wd1, namely RSGC1. The only age estimates in the literature for RSGC1 are based on the luminosities of the RSGs. Though the faintest-RSG method yields a similar age for the two clusters, the stars in RSGC1 appear to be systematically brighter than those in Wd1 \citep[][]{davies2008cool,beasor20mdot}, while RSGC1 has no known WR stars. An extensive spectroscopic study of the bright stars in RSGC1 found only early-B dwarfs, with one blue supergiant (Davies et al., in prep). This is in line with what one would expect from single-star evolutionary theory for a cluster of this age, but is in stark contrast to Wd1 which is host to many WRs and luminous hot supergiants. This suggests that stellar models {\it are} able to explain at least some clusters in this age range. Therefore, perhaps it is the ground-level assumption typically made when modelling star clusters, that of a single-age stellar population, which must be revisited in the case of Wd1. 

R136 in the LMC is another massive young cluster/association in which CSGs are known to co-exist with WRs and hot, massive post-MS stars \citep[][and references therein]{schneider2018VFTS}. Schneider et al.\ present a detailed study of around 500 hot stars in R136 and the larger surrounding association of NGC~2070, revealing a large spread of ages ranging from 1-6Myr within 18pc of the cluster core. In addition to the hot stars, there is one confirmed RSG within this zone \citep[identified by ][as Mk9 $\equiv$ W61 7-8]{britavskiy2019vlt}, which has a luminosity of $\log(L/L_\odot)=5.1$\footnote{We estimate the luminosity of W61~7-8 from its $K$-band magnitude, a distance modulus of 18.5mags, a bolometric correction of $BC_K = 2.85$ \citep{davies2018humphreys}, and an extinction of $A_K=0.19$ \citep{davies2018initial,britavskiy2019vlt}}. Across the broader 30~Dor region surrounding R136, there are dozens more cool supergiants and CSG candidates, all of which have $\log(L/L_\odot)<5.2$ \citep[][]{britavskiy2019vlt}. Hence, the simultaneous presence of hot, luminous post-MS massive stars {\it and} less luminous cool supergiants well below the H-D limit seen in Wd1 does have precedent in R136, though over a slightly larger spatial scale -- the projected spatial extent of Wd1 is around 4pc at an assumed distance of 4.12kpc. In R136/NGC~2070, this stellar diversity is easier to explain as there is compelling evidence for an extended, multi-peaked star-formation history \citep{schneider2018VFTS}. 

Therefore, from the similarity to NGC~2070, combined with the disimilarity to RSGC1, we propose that Wd1 is the product of a sustained starburst lasting several Myr. In this scenario, our RSG-based age defines the older end of the age range ($\sim$10Myr), while the eclipsing binary W13 sets the younger limit ($\sim$5Myr). With a potential age spread of 5Myr, we caution against using Wd1 as a benchmark object for calibrating stellar evolution models.

\section{Conclusions}

In this paper, we have derived bolometric luminosities for the four Red Supergiants (RSGs) and six Yellow Supergiants (YSGs) in the cluster Westerlund~1 (Wd1) using a combination of archival photometry and new mid-IR imaging with SOFIA+FORCAST. We have then compared these luminosities to the predictions from evolutionary models for a putative cluster age of 5Myr. Our findings can be summarised as follows. 

\begin{itemize}
    \item The luminosities of the RSGs and YSGs strongly disfavour an age of 5Myr, as these stars are on average 0.4dex fainter than would be expected for a cluster with this age. Formally, using an adaptation of the faintest RSG method, we find a cluster age of \rsgage, with ages younger than 6Myr excluded at the 99.9\% confidence level. 
    \item We have discussed two sources of systematic error on our luminosity estimates; photometric variability of the RSGs and YSGs, and uncertainties in the extinction law. We find that neither is able to explain the discrepancy with the expected luminosities for a 5Myr cluster. Furthermore, possible systematic errors in the `faintest RSG' age all serve to make Wd1 older, and therefore make the discrepancy worse. 
    \item In light of our results, we have re-evaluated previous age estimates for Wd1 in the literature. We argue that ages derived from stellar diversity (e.g. ratios of numbers of Wolf-Rayet stars to those of RSGs) are subject to large systematic errors and as such should be treated with caution. In addition, we have revisited the issue of isochrone fits to the pre-main sequence population of Wd1. When we apply the Gaia EDR3 distance of $4.12^{+0.66}_{-0.33}$kpc, the \citet{damineli2016extinction} extinction law, and incorporate differential extinction into the photometric errors, we find a pre-MS age of $7.2^{+1.1}_{-2.3}$Myr, though we have also shown that systematic errors could easily push this age older to within 1$\sigma$ of the `faintest RSG' age.  However, we can see no way to reconcile the tension between our results and those from eclipsing binary studies. Specifically, the upper age limit inferred from the most massive component of the W13 binary system is significantly younger than the age inferred from the RSG/YSG luminosities.
\end{itemize}

From the above evidence, we conclude that Wd1 {\it cannot} be explained by either single-star or binary stellar evolutionary models when a single age is assumed. Instead, we propose that Wd1 is the product of an extended star-forming period spanning several million years rather than an instantaneous burst. We identify the R136/NGC~2070 association as a Wd1 analogue, which contains a similarly diverse array of post-MS massive stars (Wolf-Rayets, blue/red/yellow supergiants), with RSGs of similar brightness to those in Wd1, but for which there is mounting evidence for a star-forming epoch lasting $>$5Myr.

\acknowledgments
We thank the referee for their careful reading of our manuscript, and for their suggestions which improved our paper. Based in part on observations made with the NASA/DLR Stratospheric Observatory for Infrared Astronomy (SOFIA). SOFIA is jointly operated by the Universities Space Research Association, Inc. (USRA), under NASA contract NNA17BF53C, and the Deutsches SOFIA Institut (DSI) under DLR contract 50 OK 0901 to the University of Stuttgart. Financial support for this work was provided by NASA through award \# 05 0064 issued by USRA. ERB is supported by NASA through Hubble Fellowship grant HST-HF2-51428 awarded by the Space Telescope Science Institute, which is operated by the Association of Universities for Research in Astronomy, Inc., for NASA, under contract NAS5-26555. This work makes use of the IDL software and astrolib. RDG was supported by NASA and the United States Air Force. 

%




\bibliographystyle{aasjournal}
\bibliography{sample63}{}



\end{document}